\documentclass[aip,
 amsmath,amssymb,
 reprint,%
]{revtex4-1}

\usepackage{graphicx}
\usepackage{color}
\usepackage{dcolumn}
\usepackage{bm}

\usepackage[utf8]{inputenc}
\usepackage[T1]{fontenc}
\usepackage{mathptmx}
\usepackage{etoolbox}

\makeatletter
\def\@email#1#2{%
 \endgroup
 \patchcmd{\titleblock@produce}
  {\frontmatter@RRAPformat}
  {\frontmatter@RRAPformat{\produce@RRAP{*#1\href{mailto:#2}{#2}}}\frontmatter@RRAPformat}
  {}{}
}%

\makeatother
\begin{document}

\preprint{AIP/123-QED}

\title[Minimizing Nature's Cost: Exploring Data-Free Physics-Informed Neural Network Solvers for Fluid Mechanics Applications]{Minimizing Nature's Cost: Exploring Data-Free Physics-Informed Neural Network Solvers for Fluid Mechanics Applications}
\author{Abdelrahman Elmaradny}%
 \email{aelmarad@uci.edu}
\affiliation{Department of Mechanical and Aerospace Engineering, University of California, Irvine, Irvine, CA 92697, USA
}%
\author{Ahmed Atallah}
 \affiliation{Department of Mechanical and Aerospace Engineering, University of California, San Diego, La Jolla, CA 92093, USA}

\author{Haithem Taha}
 \affiliation{Department of Mechanical and Aerospace Engineering, University of California, Irvine, Irvine, CA 92697, USA
}
%

\date{\today}

\begin{abstract}

In this paper, we present a novel approach for fluid dynamic simulations by harnessing the capabilities of Physics-Informed Neural Networks (PINNs) guided by the newly unveiled principle of minimum pressure gradient (PMPG).
In a PINN formulation, the physics problem is converted into a minimization problem (typically least squares). The PMPG asserts that for incompressible flows, the total magnitude of the pressure gradient over the domain must be minimum at every time instant, turning fluid mechanics into minimization problems, making it an excellent choice for PINNs formulation.
Following the PMPG, the proposed PINN formulation seeks to construct a neural network for the flow field that minimizes Nature's cost function for incompressible flows in contrast to traditional PINNs that minimize the residuals of the Navier-Stokes equations.
This technique eliminates the need to train a separate pressure model, thereby reducing training time and computational costs.
We demonstrate the effectiveness of this approach through a case study of inviscid flow around a cylinder, showing its ability to capture the underlying physics, while reducing computational cost and training time. 
The proposed approach outperforms the traditional PINNs approach in terms of Root Mean Square Error, training time, convergence rate, and compliance with physical metrics.
While demonstrated on a simple geometry, the methodology is extendable to more complex flow fields (e.g., Three-Dimensional, unsteady, viscous flows) within the incompressible realm, which is the region of applicability of the PMPG.

\end{abstract}

\maketitle

\section{\label{sec:Introduction}Introduction}

Simulations of fluid flows are ubiquitously needed in numerous industrial applications including aviation, marine, road vehicles, wind turbines, and environmental aerodynamics. 
Despite continuous advancements, solving fluid dynamics problems remains challenging due to the complexity of the governing equations, especially for turbulent flows at high Reynolds numbers. Traditional methods, such as analytic solutions and computational fluid dynamics (CFD) simulations, have limitations in handling real-world complexities efficiently.

The emergence of Physics-Informed Neural Networks (PINNs) offers a promising avenue to address these challenges \cite{raissi_physics-informed_2019}. 
Although this technology has not yet matured enough to outperform traditional techniques, it may be prudent to invest in the development of this novel approach. 
By integrating the laws of physics directly into neural network learning, PINNs ensure that solutions adhere to underlying physical principles, not solely relying on data-driven approaches. 
PINNs aim to minimize residuals in governing equations, boundary and initial conditions, and deviation from experimental measurements, effectively converting physics problems into least-squares minimization problems. 
However, minimizing residuals may lead to additional spurious solutions \cite{seliger1968}.

Currently, the Principle of Minimum Pressure Gradient (PMPG) \cite{taha2022does,taha2023} provides a natural cost function for incompressible flows. 
It asserts that an incompressible flow evolves from one time instant to another in order to minimize the total pressure gradient over the domain. 
It was shown that Navier-Stokes equations represent the first-order necessary condition for minimizing the pressure gradient cost.
This principle transforms fluid mechanics problems into minimization problems, facilitating the determination of flow parameters.

PINNs and PMPG share a profound connection in modeling incompressible fluid mechanics. 
On the one hand, the PMPG provides a suitable physical cost function for PINNs, on the other hand, the PINN formulation offers a convenient framework for applying PMPG to large-scale problems where flow field parametrization may be challenging.
The use of PMPG as Nature's cost function, applied to Physics-Informed Neural Networks was proposed by the authors \cite{atallah2024novel} and independently by Alhussein and Daqaq \cite{alhussein2024principle} where they demonstrated its efficiency in solving the Lid-Driven Cavity problem. 
In this work, we provide a detailed presentation and analysis of this novel PMPG-based PINN formulation, demonstrating its success in solving inviscid incompressible flow over a two-dimensional cylinder without relying on training data. Extension to unsteady, viscous, three-dimensional problems will be addressed in future work.
In this work, we also investigate the use of vorticity transport equation as a side constraint on the minimization problem. By taking the curl of the momentum equation, we avoid training an extra pressure model, reducing the computational cost significantly. 
In other words, setting the curl of the pressure gradient to zero while minimizing the pressure gradient cost over a network that naturally satisfies the continuity equation is an alternate and more efficient way to impose the conservation of momentum than minimizing the residuals of the momentum and continuity equations.

In terms of applications, aside from being a promising Data-Free solver, our approach shows potential in filtering noisy data within incompressible flow fields, effectively correcting deviations from the original field while adhering to physics principles. 
By incorporating conservation laws (represented by the PMPG) and boundary conditions during training, noisy flow field data can be effectively filtered.
This underscores the potential of PMPG-based PINNs to integrate physical knowledge into neural network-based modeling for noise filtration.

The paper is structured as follows: Section \ref{sec:PMPG} provides an introduction to solving fluid mechanics problems as minimization problems using the PMPG.
Section \ref{sec:PINNs} presents the basics of Physics-Informed Neural Networks (PINNs). 
Section \ref{sec:PMPG based PINNs} represents the core of the work, highlighting the framework of integrating PINNs and the Principle of Minimum Pressure Gradient (PMPG) into a data-free solver for incompressible flows. 
Section \ref{sec:Results} then provides an analysis of the application of our approach to inviscid flow over a two-dimensional cylinder. 
Section \ref{sec:Comparison} presents a comparison between different techniques with our approach to gain a better understanding of the capabilities and limitations of the proposed approach. 
Section \ref{sec:Numerics} provides an analysis of the effect of changing the parameters of the neural network and the sensitivity of the solution with respect to the number of neurons and the weighting coefficients of the cost function.
In Section \ref{sec:Physics-Driven regularization}, the significance of imposing the curl of the momentum equation as a side constraint is discussed, showing its regularizing effect on training the Neural Network. Finally, the paper is concluded with a summary of findings in Section \ref{sec:Conclusions}.

\section{\label{sec:PMPG}Fluid mechanics as a minimization problem}

Fluid flows adhere to the fundamental principles of mass and momentum conservation. 
The continuity equation symbolizes mass conservation, defined for incompressible flows as a divergence-free constraint on the velocity field :
\begin{equation} 
\nabla \cdot \mathbf{u} = 0,
\label{eqn1}
\end{equation}
where $\mathbf{u}$ denotes the velocity field vector.

The Navier-Stokes equations govern momentum conservation, representing a system of three coupled partial differential equations that describe the evolution of a fluid's velocity field over time. These equations in their simple form are expressed as :
\begin{equation}
\rho \Big ( \frac{\partial \mathbf{u}}{\partial t} + \mathbf{u} \cdot \nabla \mathbf{u} \Big )= -\nabla P + \nabla \cdot \mathbf{\tau}  ,
\end{equation}
where $P$ signifies pressure, $\rho$ is the fluid density, $\nabla \cdot \mathbf{\tau}$ represents viscous forces and the term between brackets in the left-hand side of the equation is the acceleration of the flow.

The solution to these equations, subject to suitable boundary and initial conditions such as no-penetration and no-slip conditions, provides the velocity and pressure fields.
However, solving these equations typically necessitates advanced mathematical and computational techniques due to their nonlinear and complex nature, especially in turbulent scenarios.

In this study, we embrace the Principle of Minimum Pressure Gradient (PMPG), departing from traditional methods that rely on Newtonian mechanics to derive the Navier-Stokes equations. Instead, we advocate for variational principles to describe fluid dynamics.
The use of variational methods has been largely confined to ideal fluids under Hamilton's principle of least action \cite{seliger1968,bretherton1970,salmon1988,morrison1998}.
Other notable efforts managed to extend the principle of least action to account for the dissipative viscous terms in Navier-Stokes \cite{fukagawa2012variational,galley2014principle,gay2018lagrangian,sanders2024canonical}.
However, the statement of the principle of least action presents only a stationary functional that is not 
necessarily minimum.
Therefore, although elegant, these extensions cannot be used in a minimization formulation, in contrast to the PMPG.

Recent advancements in fluid analysis, leveraging Gauss' principle of least constraint \cite{papastavridis2014}, as demonstrated by Taha et al. \cite{taha2022does,taha2023variational,taha2023}, offer a pure minimization formulation of incompressible fluid mechanics \cite{gonzalez2022variational,taha2022flow,taha2023refining,shorbagy2024magnus}, in comparison to prior variational methods.

In analytical mechanics, forces are typically categorized as impressed or driving forces $\mathbf{F}$ and constraint forces $\mathbf{R}$. 
The equation of motion is thus written as:
\begin{equation}
m \mathbf{a}= \mathbf{F} + \mathbf{R},
\end{equation}
where $\mathbf{a}$ represents inertial acceleration. Constraint forces maintain geometric or kinematic constraints and are passive. 
Examples include the tension in a pendulum's string or normal forces on a surface. Gauss' principle stipulates that Nature minimizes the magnitude of constrained forces $\mathbf{R}$, hence the name least constraint. The Gaussian cost was explicitly written by Jacobi as:
\begin{equation}
Z=  \frac{1}{2}\sum_{}^{} m\Big(\mathbf{a}-\frac{\mathbf{F}}{m}\Big)^2
\label{eqn4}
\end{equation}
$Z$ must be minimized at every instant, distinguishing it from the stationary principle of least action \cite{taha2023}.

It is well established in the mathematical theory of incompressible flows that pressure is the Lagrange multiplier that ensures the continuity constraint(e.g., \cite{moin1980numerical,gresho1987pressure,chorin1990mathematical,kambe2009geometrical,arnold1966geometrie}).
That is, the pressure force is a constraint force, and its role is to ensure the continuity constraint.
Consequently, Taha et al. \cite {taha2022does,taha2023} demonstrated that, for an incompressible flow without an external pressure gradient, the pressure force is workless, as known from the Helmholtz orthogonal projection.
Hence, applying Gauss' principle to fluid mechanics, where the inertial acceleration is written as $\mathbf{a}$ $=$ $\frac{\partial \mathbf{u}}{\partial t} + (\mathbf{u} \cdot \nabla) \mathbf{u}$ and the impressed force is the viscous term, Taha et al. \cite{taha2022does,taha2023} proposed minimizing the following cost function:
\begin{equation}
\label{eq:appellian_cost}
    \mathcal{A} = \frac{1}{2}\int_{\Omega}^{} \rho \Big(\frac{\partial \mathbf{u}}{\partial t} + (\mathbf{u} \cdot \nabla) \mathbf{u} - \nu \nabla^2 \mathbf{u}\Big)^2 d\mathbf{x},
\end{equation}
subject to the continuity constraint (Equation \eqref{eqn1}) and the no-penetration boundary condition:
\begin{equation}
\begin{aligned}
    \mathbf{u} \cdot \mathbf{n}=0, \quad \text{on} \ \delta\Omega.
\end{aligned}
\end{equation}
The no-slip boundary condition can be imposed in viscous flows, though it's not necessary for the minimality of the cost (Equation \eqref{eq:appellian_cost}).
Note that the cost $\mathcal{A}$ is nothing but the total pressure gradient over the domain.

Derived from fundamental principles of analytical mechanics, this principle seamlessly transforms fluid mechanics problems into minimization problems.

\section{\label{sec:PINNs}Physics-informed Neural Networks}
Neural networks have made significant advances in several fields, proving to be a powerful tool in processing and analyzing large complex datasets.
For instance, in computer vision, they have enabled significant advancements in image recognition and object detection \cite{krizhevsky2012imagenet,he2016deep,redmon2016you}. 
Recommendation systems leverage neural networks to personalize user experiences by analyzing preferences and patterns in user behavior \cite{covington2016deep,he2017neural}. 
In medical diagnosis, neural networks aid in the interpretation of medical images, such as X-rays and Magnetic Resonance Imaging (MRIs), and assist in disease detection and prognosis \cite{esteva2017dermatologist,shen2017deep}. 
Neural networks are also widely used in forecasting tasks, including financial market prediction \cite{tsantekidis2017forecasting,ding2015deep} and weather forecasting \cite{shi2015convolutional,lipton2015learning}. 
In game theory, they are employed to model complex interactions and strategies, enabling the development of intelligent agents \cite{silver2016mastering,lanctot2017unified}. 
Cognitive science benefits from neural networks in modeling brain processes and understanding human cognition \cite{mnih2015human,hassabis2017neuroscience}. 
Neural networks are also used in genomics to analyze DNA sequences, predict gene functions, and identify genetic markers associated with diseases \cite{angermueller2016deep,zhou2015predicting}. 
In the realm of autonomous driving, they enabled vehicles to perceive their environment, make decisions, and navigate safely on the road \cite{bojarski2016end,chen2015deepdriving}. 
In the entertainment side, neural networks are employed in playing board and computer games, achieving superhuman performance in games like Go and chess \cite{silver2017mastering}.
Furthermore, neural networks play a crucial role in speech recognition and natural language understanding \cite{hinton2012deep,bahdanau2014neural,vaswani2017attention}. 
The recent surge in Large Language Models (LLMs) over the past two years, where Neural Networks architecture is employed in training \cite{wu2023brief,roumeliotis2023chatgpt}, has reaffirmed their position as a leading artificial intelligence technique, propelling human advancement. Overall, the widespread integration of neural networks across diverse domains underscores their adaptability and efficacy in addressing a myriad of complex challenges.


The first employment of Neural Networks to solve Ordinary Differential Equations (ODEs) and Partial Differential Equations (PDEs) dates back to 1989. They were treated as suitable function approximators that satisfy boundary and initial conditions in a constrained minimization approach \cite{lagaris1998artificial,lagaris2000neural}. The efforts of Owhadi were pioneering in applying physical knowledge about the solution in the training \cite{owhadi2015bayesian}. Hence, the learning process becomes physics-informed.
The real introduction of the term Physics-Informed Neural Networks (PINNs) was not until 2017 in the seminal papers of Raissi et al.\cite{raissi2017physics,raissi2017physics2,raissi_physics-informed_2019} solving non-linear PDEs.
This pioneering work marked a significant milestone in the application of PINNs for solving non-linear PDEs. Raissi et al. investigated the effectiveness of PINNs in tackling a range of challenging PDEs, including the Schrödinger equation, Allen–Cahn equation, Navier–Stokes equation, and Korteweg–de Vries equation. 
By incorporating physics-based constraints into the neural network architecture, PINNs demonstrated their capability to effectively capture the complex dynamics and nonlinearities inherent in those PDEs. 
The study by Raissi et al. demonstrated the broad applicability and promising potential of PINNs in the realm of non-linear PDEs, opening up new avenues for solving complex physical problems with data-driven neural network models. 
Subsequent efforts applied the concepts of PINNs to heat transfer \cite{cai_physics-informed_2021} and high-speed flows \cite{mao_physics-informed_2020}. 
In their work, Blechschmidt et al. \cite{blechschmidt_three_2021} explored the application of deep learning architectures, specifically convolutional neural networks (CNNs), for solving PDEs. 
Their study demonstrated the effectiveness of deep learning models in accurately approximating the solutions of PDEs and showed their potential for capturing intricate physics phenomena.

Currently, Physics-Informed Neural Networks (PINNs) stand at the forefront of computational science, blending deep learning with physics-based modeling.
Central to the concept of PINNs is the direct integration of physical laws, typically represented as differential equations, into the architecture of neural networks and its cost function.
This integration ensures that predictions made by the network are not solely reliant on data but also align with the fundamental physical principles governing the system.
This alignment is achieved by introducing a loss term, known as the physics-informed loss, during the network's training process, which quantifies the residuals of the governing equations when evaluated at the model's outputs.

Mathematically, the PINN model is expressed as:
\begin{equation}
\mathbf{y}(\mathbf{x}, t; \boldsymbol{\theta}) = F_{NN}(\mathbf{x}, t; \boldsymbol{\theta}).
\end{equation}
Here, $\mathbf{y}$ denotes the model's output, $\mathbf{x}$ represents the input, $t$ signifies time, and $\boldsymbol{\theta}$ encompasses the parameters of the neural network $F_{NN}$.
These parameters $\boldsymbol{\theta}$ are determined through the network's training process, which is achieved by minimizing the total loss function $\mathcal{L}$:
\begin{equation}
    \mathcal{L} = \mathcal{L}_{data} + \lambda \mathcal{L}_{physics},
\end{equation}
where $\mathcal{L}$ comprises a data loss term $\mathcal{L}_{data}$ and a physics loss term $\mathcal{L}_{physics}$, with a weighting coefficient $\lambda$ that controls the relative importance of physical knowledge to the PINNs model in comparison to the data.
This cost function ensures that the PINN learns from the available data while simultaneously adhering to the physical constraints imposed by the governing equations.


By minimizing the PINN loss function, the neural network parameters are optimized to simultaneously fit the available data and satisfy the underlying physics, making PINNs a powerful tool for combining data-driven and physics-based approaches in various scientific and engineering domains. 
This advantage makes PINNs particularly suited for reconstructing fluid flows from sparse or noisy data\cite{liu_multi-fidelity_2019}.

\section{\label{sec:PMPG based PINNs}PINNs driven by the Principle of Minimum Pressure Gradient (PMPG)}

With physical principles and recent advancements in Neural Networks, the exploration of their utility in addressing physics problems has sparked interest in solving fluid mechanics problems without relying on external data, giving rise to "Data-Free" Physics Constrained Neural Networks. 
In this emerging field of research, the Neural Networks are trained, not to match given data, but to adhere to physical governing equations.
Unlike conventional methods that hinge on partial differential equations and data-driven optimization, this variant of Physics-Informed Neural Networks (PINNs) delves into the minimization of physical quantities as substitutes for data-driven training inputs. Notably, the efforts of Nguyen et al. \cite{nguyen-thanh_deep_2020} and Goswami \cite{goswami_transfer_2020} have demonstrated the efficacy of this approach in structural dynamics and brittle fracture mechanics, respectively. 
Yet, a critical question remained: could a similar framework be adapted for fluid mechanics problems?

In the standard procedure adopted by Data-Free Neural Networks, a neural network for the target function \( \mathbf{y}(\mathbf{x}, t; \boldsymbol{\theta}) \) is initially constructed. Following this, boundary conditions and initial conditions specific to the problem are satisfied. Subsequently, derivatives of the target function are computed. 
Finally, a minimization quantity \( \mathbf{E} \) specific to the physical problem at hand is integrated into the loss function. 
Typically, this quantity is a function of the \( \mathbf{y}(\mathbf{x}, t; \boldsymbol{\theta}) \) and its derivatives, making it inherently a function of Neural Network parameters \( \mathbf{E} = F_{NN}(\mathbf{y},\frac{\partial \mathbf{y}}{\partial x_i}, \frac{\partial \mathbf{y}^2}{\partial x_i \partial x_j}, \frac{\partial \mathbf{y}}{\partial t} ....; \boldsymbol{\theta}) \), thereby guiding the network towards learning and optimizing its performance based on the physical principles of the problem. 
The key task here is to identify the fundamental physical quantity that must be minimized at the correct solution.
In the standard approach of PINNs, a weighted sum of the residuals of the governing equations is typically considered as the cost function to be minimized. 
In this work, we propose a more natural cost for incompressible flows: the total magnitude of the pressure gradient over the domain.
According to the Principle of Minimum Pressure Gradient (PMPG) \cite{taha2022does,taha2023}, this is the fundamental quantity that Nature minimizes at every instant of time for any incompressible flow.


The integration of the Principle of Minimum Pressure Gradient (PMPG) with Physics-Informed Neural Networks (PINNs) (termed PMPG-based PINNs) combines the strengths of both methodologies to tackle fluid dynamics problems more effectively, utilizing a nature-inspired cost function. 
In this approach, we reformulate the Navier-Stokes equations in terms of the velocity field and its derivatives, effectively avoiding the explicit calculation of the pressure field. 
The network's architecture and training process remain similar to standard PINNs, but the loss function is modified to include the PMPG cost: 
\begin{eqnarray}
\mathcal{L}_{PMPG} =\sum \left| \frac{\partial \mathbf{u}}{\partial t} + (\mathbf{u} \cdot \nabla) \mathbf{u} - \nu \nabla^2 \mathbf{u}
\right|^2 .
\label{loss function}
\end{eqnarray}

For brevity and clarity, the notation $(\mathbf{x}, t; \boldsymbol{\theta})$, which typically indicates spatial coordinates, time, and neural network parameters, has been omitted from the terms in Eq. \ref{loss function}. 
In this formulation, $\mathbf{u}(\mathbf{x}, t; \boldsymbol{\theta})$ represents the predicted velocity field output by the neural network, parametrized by weights and biases of the neural network. 
The PMPG loss includes the squared norm of the pressure gradient, aiming to minimize it over the flow domain in accordance with the PMPG.

The incorporation of PMPG into PINNs offers several advantages. 
First, it streamlines the optimization problem, reducing computational complexity, by focusing on Nature's cost function in comparison to a \textit{devised} cost function using Least Squares. Second, unlike classical PINNs that often employ separate neural networks for modeling velocity and pressure fields, a PMPG-based PINN needs only to train the velocity field and its derivatives; no network is constructed for pressure.
This approach offers an efficient and less computationally demanding method for modeling complex fluid dynamics.
Third, in scenarios where multiple solutions exit, this approach will converge to the most physical one from the perspective of Gauss' principle of least constraint.

\begin{figure*}[ht]
    \centering
    \includegraphics[width=.8\textwidth]{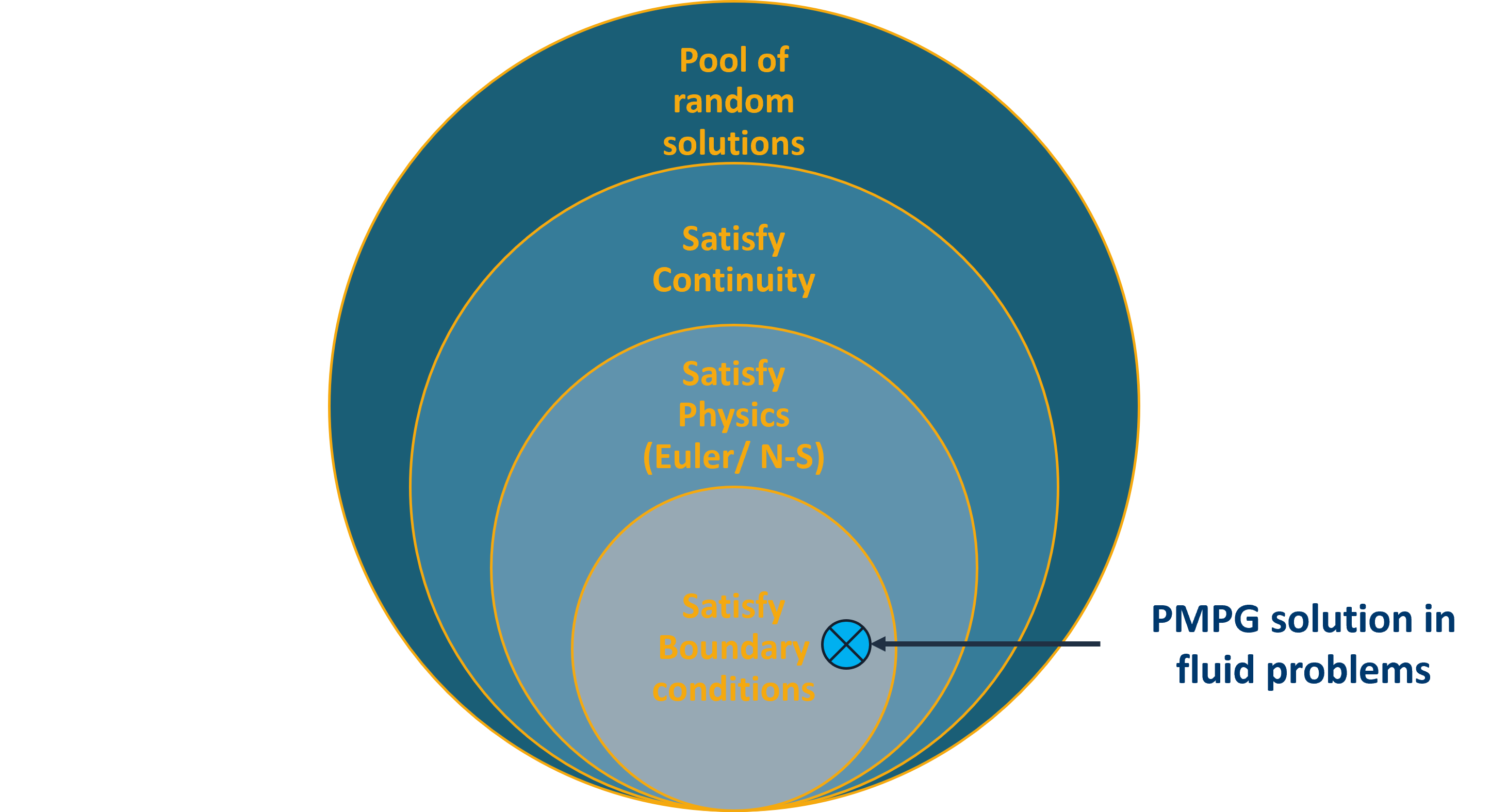}
    \caption{A Venn diagram for how the Neural Network is envisioned to search for the optimal model narrowing down the possible solutions based on boundary and physics constraints and then minimizing the Pressure Gradient to find the correct solution.}
    \label{fig:venn diagram}
\end{figure*}

By incorporating a combination of loss functions that encompass boundary conditions and other physical constraints, the network is directed towards achieving the PMPG optimal solution [Figure \ref{fig:venn diagram}]. 
This process effectively narrows down the infinite solution space to a more constrained set through successive layers of constraints.
Our methodology entails enforcing adherence to governing conservation laws, including mass conservation, and geometric boundary conditions. 
Subsequently, we pursue pressure gradient minimization to attain convergence to the solution, devoid of external data reliance. 
Leveraging the capabilities of neural networks and autodifferentiation for computing derivatives, this approach exhibits promise in tackling fluid mechanics problems in a data-independent manner. 
Furthermore, this approach circumvents the necessity for pressure field computations, thereby mitigating complications associated with pressure-velocity coupling inherent in traditional Navier-Stokes  solvers.

The optimization problem can be posed as follows: the function to be minimized is the Pressure Gradient squared integrated over the domain, subject to constraints of continuity and equilibrium while satisfying boundary conditions. 
In the case of steady incompressible flows, this problem can be expressed mathematically as:
\begin{equation}
\begin{aligned}
\label{eq:steady_problem_formulation}
\min_{\mathbf{u}(\mathbf{x}:\boldsymbol{\theta})} \quad & \mathcal{A}= \frac{1}{2}\int_{\Omega} \rho \Big((\mathbf{u} \cdot \nabla) \mathbf{u} - \nu \nabla^2 \mathbf{u}\Big)^2 d\mathbf{x}\\
\textrm{s.t.} \quad & \nabla \times \Big((\mathbf{u} \cdot \nabla) \mathbf{u} - \nu \nabla^2 \mathbf{u}\Big) = 0, \\
& \nabla \cdot \mathbf{u} =0,\\
& \mathbf{u} \cdot \mathbf{n}=0, \quad \text{on} \ \delta\Omega, \\
& (\mathbf{u} - \mathbf{u}_{wall}) \cdot \mathbf{t} = 0, \quad \text{on} \ \delta\Omega
\end{aligned}
\end{equation}
where $\mathbf{n}$, $\mathbf{t}$ are normal and tangential vectors to the boundary $\delta\Omega$.

In this work, we use the penalty method [Section \ref{secpenaltymethod}] to tackle the constraints of equilibrium and boundary conditions. 
However, the continuity constraint is automatically satisfied by making use of the streamfunction formulation.
In other words, we construct a neural network for the streamfunction $\psi$ and use automatic differentiation to obtain the velocity field $\mathbf{u}$ instead of constructing a neural network for the velocity field directly.
In this way, the continuity constraint is automatically satisfied.

\subsection{Equilibrium Constraint}

To enforce conservation of momentum without constructing a separate pressure model, we take the curl of momentum equation i.e, the vorticity transport equation in its non-conservative form: 
\begin{equation}
\label{eq:vorticity_transport}
    \nabla \times \Big((\mathbf{u} \cdot \nabla) \mathbf{u} - \nu \nabla^2 \mathbf{u}\Big) = 0.
\end{equation}
This condition is imposed as a hard constraint through the penalty method details in Section \ref{secpenaltymethod}.


\subsection{Penalty Method}
\label{secpenaltymethod}

Addressing the intricate trade-offs between various cost functions is a common challenge encountered in problems characterized by multi-objective optimization.
The resolution of conflicts among these objectives is largely dependent upon the weights $\lambda$ assigned to each term.
Certain terms, such as boundary conditions and conservation laws, mandate full compliance.

Among the several techniques that can be used to enforce hard constraints (e.g., the augmented Lagrangian method \cite{lu2021physics}), the penalty method is widely employed for imposing hard constraints during neural network training due to its seamless integration into the loss function. 

Mathematically, the penalty method introduces a penalty term into the loss function to quantify the degree of constraint violation. 
This penalty term typically increases with larger violations, forcing the neural network to search in a direction that decreases the constraint violation \cite{luenberger1984linear,long2018pde}.

Consider a neural network model characterized by an input vector \(\mathbf{x}\) and parameters \(\boldsymbol{\theta}\), where the cost function is given by \(\mathcal{L}(\mathbf{x};\boldsymbol{\theta})\).
This cost function needs to be minimized while satisfying an equality constraint function  \(\mathcal{H}(\mathbf{x};\boldsymbol{\theta})\). 
So the constrained minimization problem is defined as:
\begin{equation}
\begin{aligned}
\label{constrainedopt}
\min_{\mathbf{\theta}} \quad & \mathcal{L}(\mathbf{x};\boldsymbol{\theta}) \\
\textrm{s.t.} \quad & \mathcal{H}(\mathbf{x};\boldsymbol{\theta})=0.
\end{aligned}
\end{equation}
The penalty method appends the loss function with a penalty term that represents the constraint violation: 
\begin{equation}
\mathcal{L}_{\text{Total}}(\mathbf{x};\boldsymbol{\theta},C) = \mathcal{L}(\mathbf{x};\boldsymbol{\theta}) + C  \mathcal{P}(\mathcal{H}(\mathbf{x};\boldsymbol{\theta})),
\end{equation}
where \({\text{C}}\) denotes the penalty coefficient regulating the significance of the constraint, while \(\mathcal{P}(\cdot)\) represents a penalty function that quantifies the extent of constraint violation. 
The selection of the penalty function \(\mathcal{P}(\cdot)\) depends on some factors. 
It should be continuous, positive and, equal to zero when the constraint is satisfied. 
In this case, a quadratic penalty function is used:
\begin{equation}
\mathcal{P}(\mathcal{H}(\mathbf{x};\boldsymbol{\theta}))= \mathcal{H}(\mathbf{x};\boldsymbol{\theta})^2.
\end{equation}

The main idea of using the penalty method is to convert the constrained minimization problem (Equation \eqref{constrainedopt}) to an unconstrained minimization by augmenting the cost function with the penalty term \(\mathcal{P}(\cdot)\):

\begin{equation}
\begin{aligned}
\min_{\mathbf{\theta}} \quad & \mathcal{L}(\mathbf{x};\boldsymbol{\theta}) + C  \mathcal{P}(\mathcal{H}(\mathbf{x};\boldsymbol{\theta})).
\end{aligned}
\end{equation}
For instance, when imposing some equality constraint $\mathcal{H}(\theta)=0$ to align with a feasible tolerance of \(\mathcal{\epsilon}\), a quadratic penalty function may be employed. In this scenario, the modified loss function takes the form:

\begin{equation}
\begin{aligned}
\min_{\mathbf{\theta}} \quad &  \mathcal{L}_{\text{Total}}(\mathbf{x};\boldsymbol{\theta},C)=\mathcal{L} (\mathbf{x};\boldsymbol{\theta}) + C  \left( \mathcal{H}(\mathbf{x};\boldsymbol{\theta}) - \epsilon\right)^2,
\end{aligned}
\end{equation}
where \(\mathcal{\epsilon}\) denotes the acceptable tolerance of the constraint's violation.

Throughout the training process, the neural network seeks to minimize the modified loss function \(\mathcal{L}_{\text{Total}}(\mathbf{x};\boldsymbol{\theta})\). 
By incorporating the penalty term, the network is incentivized to identify parameter values that optimize the objective while concurrently adhering to the imposed constraint. 
The penalty coefficient \({\text{C}}\) must be large enough to ensure proper satisfaction of the constraint.
In our current model, \({\text{C}}\) was set to $10^{6}$, with \(\mathcal{\epsilon}\) = $0.01$.

Considering all the aforementioned factors, the total loss function used in this research, based on the formulation mentioned in Equation \ref{eq:steady_problem_formulation}, is expressed as:
\begin{multline}
\label{eq:PMPG + curl}
\mathcal{L}(\mathbf{x};\boldsymbol{\theta}) = \lambda_1 \mathcal{L}_{\text{BC}}(\mathbf{x};\boldsymbol{\theta})\\+
\lambda_2 \mathcal{L}_{\text{Eqm}}(\mathbf{x};\boldsymbol{\theta}) + 
\lambda_3\mathcal{L}_{PMPG}(\mathbf{x};\boldsymbol{\theta}).
\end{multline}
Here, $\mathcal{L}_{\text{BC}}$, $\mathcal{L}_{PMPG}$ and $\mathcal{L}_{\text{Eqm}}$ represent the losses associated with boundary conditions, equilibrium (Equation \ref{eq:vorticity_transport}), and the PMPG, respectively. 
The corresponding weights are $\lambda_1$, $\lambda_2$, $\lambda_3$. 

The penalty method is employed to enforce boundary conditions and conservation of momentum. 
While the conservation of mass can also be enforced by hard constraining the divergence of the velocity field to be zero, we opt to have this important constraint naturally satisfied by the structure of the neural network.
Since the continuity constraint is linear, one can construct the neural network to automatically satisfy it \cite{hendriks2020linearly}.
In the cases of two-dimensional flows it can be simply achieved by constructing a Neural Network for the streamfunction $\psi$ instead of the velocity field.
This approach can also be generalized for three-dimensional flows using the concept of vector potential.
That is, we will construct a Neural Network for the vector potential $\mathbf{A}$ from which the velocity field can be obtained as $\mathbf{u}= \nabla \times \mathbf{A}$.
As such, the continuity constraint $\nabla \cdot \mathbf{u}=0$ will automatically be satisfied.

In all our simulations, the RMS error of boundary conditions was $0.7 \%$, with a maximum error of $1.1 \%$, indicating the efficacy and precision of the penalty method's application.
While further refining of the accuracy is conceivable through a decreased tolerance value \(\mathcal{\epsilon}\), the resulting error magnitudes were deemed satisfactory for our purposes.



\section{\label{sec:Results}Proof of Concept}

In this paper, we were mainly interested in providing a proof-of-concept of the proposed new approach of PMPG-based PINNs, analyzing its strengths and limitations in comparison to traditional PINNs.
This proof-of-concept is demonstrated on a simple classical example in incompressible fluid mechanics: the steady ideal flow over a two-dimensional cylinder.
Due to its simplicity, the example provides insights into the capabilities and limitations of the proposed approach, avoiding a cluttered picture due to a complex flow field.
Notably, this example possesses an exact analytical solution, as represented by Eq. (\ref{eq:theoretical equation}), facilitating a rigorous comparison with our obtained results. 
The analytical solution for the streamfunction $\psi$ is expressed as:
\begin{equation}
\label{eq:theoretical equation}
    \psi(r,\theta) = U_\infty \Big(r-\frac{a^2}{r}\Big)\sin\theta.
\end{equation}

Given the two-dimensional nature of the problem, we formulate our output as the streamfunction $\psi$ to inherently satisfy the conservation of mass. 
Subsequently, differentiation of the streamfunction $\psi$ yields the velocity components in the $x$ and $y$ directions:

\begin{equation}
    u_x = \frac{\partial\psi}{\partial y}, \qquad u_y = -\frac{\partial\psi}{\partial x}.
\end{equation}

\subsection{\label{sec:Problem formulation}Problem formulation}

\begin{figure}[h!]
\centering
\includegraphics[width=0.45\textwidth]{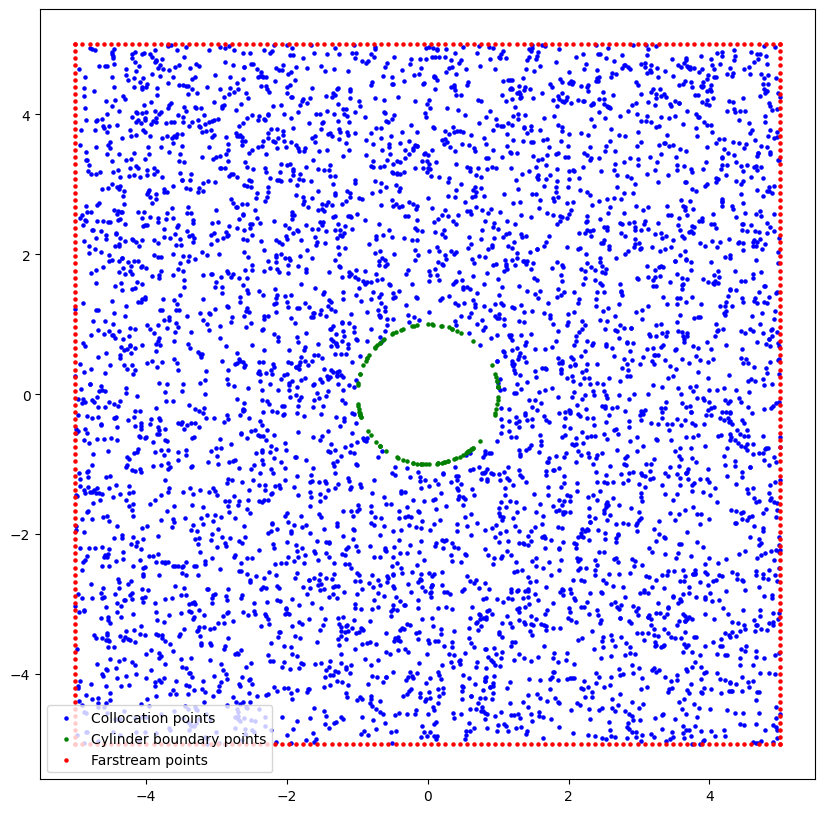}
\caption{\label{fig:Mesh grid} Map of training points: interior points in blue, cylinder boundary points in green, and far-field points in red.}
\end{figure}

The selected problem domain encompasses a two-dimensional cylinder with a unity radius $a$, surrounded by a collection of 5000 randomly distributed points within a square whose side length is $10 a$.
100 uniformly distributed points were considered on the cylinder circumference to enforce the No-Penetration boundary condition via the penalty method discussed in Section \ref{secpenaltymethod}.
Specifically, the radial velocity was constrained to zero at the boundary points. 
Additionally, 100 points were generated on each edge of the outer boundary to enforce the far-field boundary condition $(u_x,u_y)=(1,0)$.
Figure \ref{fig:Mesh grid} shows the selected domain with the interior and boundary points.

\subsection{\label{sec:NN structure}Neural Network structure}

\begin{figure}[!]
\centering
\includegraphics[width=0.4\textwidth]{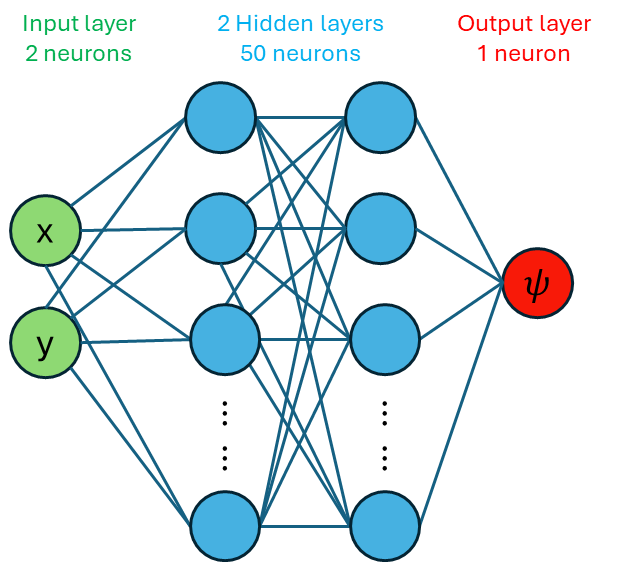}
\caption{\label{fig:NN structure} The Neural Network structure is comprised of two hidden layers of 50 neurons each.}
\end{figure}

Our Neural Network structure is a Feed-Forward Fully-Connected Neural Network comprising two hidden layers, each composed of 50 neurons. 
The input layer consists of two neurons representing the spatial coordinates $x$ and $y$, while the output layer comprises a single neuron representing the streamfunction $\psi$, as illustrated in Figure \ref{fig:NN structure}.
The selected activation function is the hyperbolic tangent (\textit{Tanh}) function. Training is performed using the \textit{Adam} optimization algorithm - a type of stochastic optimization algorithm that is commonly used for training deep neural networks \cite{kingma2014adam}.
It should be noted that modifying the number of neurons and the number of hidden layers can be pursued to enhance the accuracy of the model in the case of more complex flow fields \cite{mo2022data}.
In Section \ref{subsec:neoruns}, the effect of the number of neurons is studied.

\subsection{\label{sec:error analysis}Error analysis}

\begin{figure*}[!]
\centering
\includegraphics[width=0.9\textwidth]{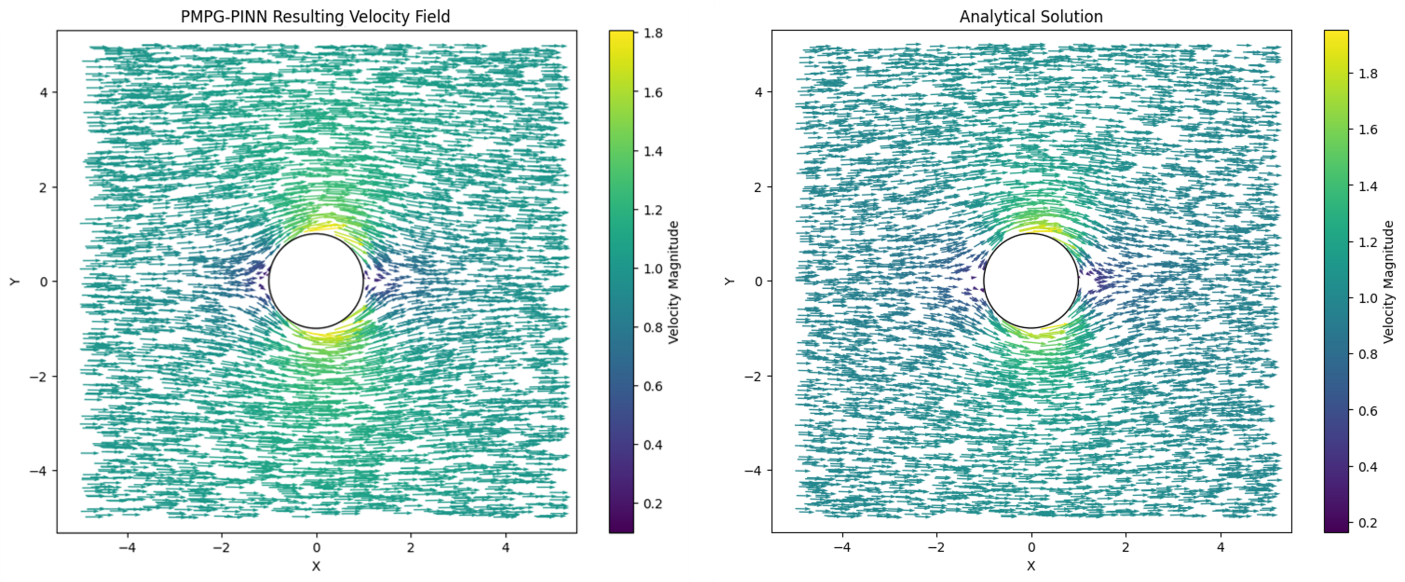}
\caption{\label{fig:Velocity field} PMPG-PINN Resulting velocity field vs Analytical solution velocity field.}
\end{figure*}

\begin{figure}[!]
\centering
\includegraphics[width=0.4\textwidth]{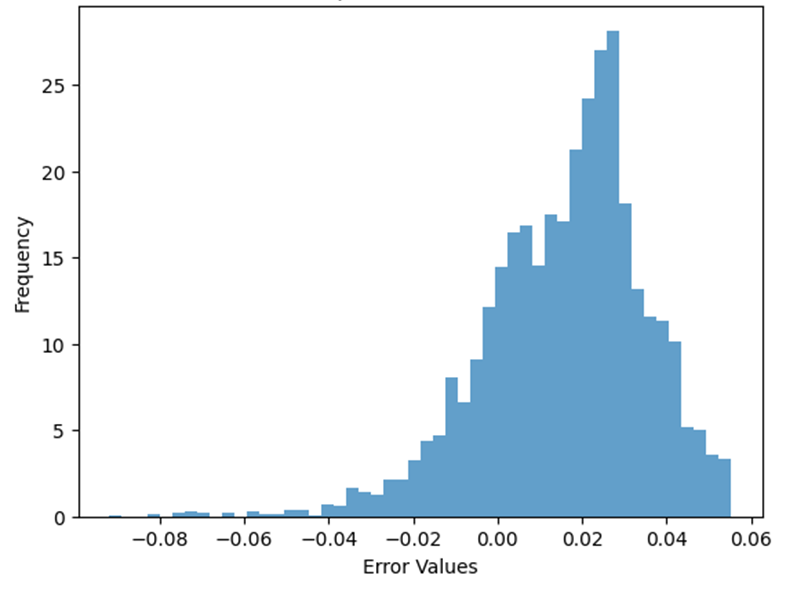}
\caption{\label{fig:Error histogram} Velocity error histogram.}
\end{figure}

After training the model for 30000 epochs, the resulting streamfunction $\psi$ was differentiated to obtain the velocity field.
Visual comparison with the theoretical velocity field reveals a remarkable agreement as shown in Figure \ref{fig:Velocity field}.
The Root Mean-Squared Error (RMSE) in the velocity field compared to the analytical solution at the collocation points was found to be 1.8\%. The error histogram, depicted in Figure \ref{fig:Error histogram}, exhibits a standard deviation of 1.9\% and a skewness of -0.7889.

\subsection{\label{sec:Physics checks}Physical measures}

In order to measure to what extent our results are physical, several metrics were calculated and plotted. 

\paragraph{The divergence of the velocity vector} provides an indication of the compatibility of the generated velocity field with the conservation of mass. 
When the Root Mean-Square of $\nabla \cdot \mathbf{u}$ is calculated for the resulting flow field, it yielded a negligible value of $4.8 \times 10^{-8}$. 
This result is expected due to the use of the streamfunction formulation, which ensures an inherent satisfaction of continuity.

\begin{figure}[!]
\centering
\includegraphics[width=0.4\textwidth]{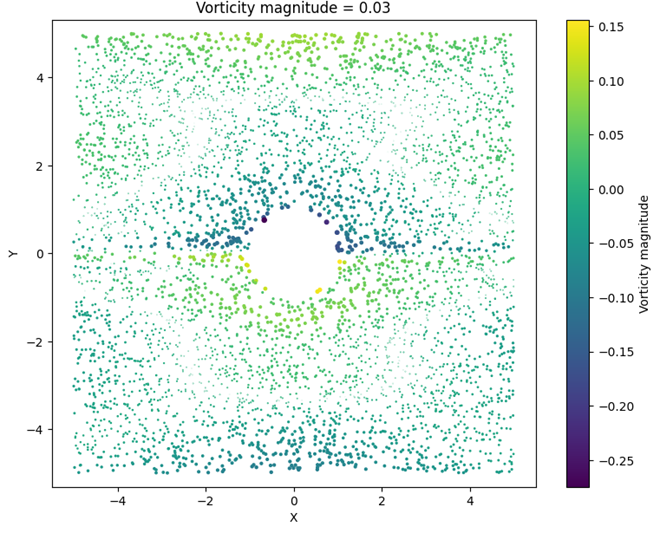}
\caption{\label{fig: Vorticity map}  Vorticity Contour of the PMPG-PINN resulting field.}
\end{figure}

\paragraph{The curl of the velocity vector} offers a measure of the vorticity of the flow.
It should be noted that we have not imposed any explicit constraints on vorticity during training. 
Yet, the resulting vorticity field naturally came out as irrotational, as shown in Figure \ref{fig: Vorticity map}.
The vorticity RMS was found to be 0.03, demonstrating good agreement with the theoretical equations. 
It may be important to emphasize that the resulting flow field is not only irrotational but also symmetric with zero circulation. 
Note that an irrotational flow with an arbitrary value of circulation is a perfectly legitimate mathematical solution of Euler's equation in the cylinder problem.
Hence, the corresponding residuals in momentum, continuity, no-penetration boundary condition, and far-field boundary condition are identically zero for any value of circulation.
Therefore, a traditional PINN is not immune to converging to one of these solutions.
In contrast, the proposed PMPG-PINN approach will converge to only one solution among this family of solutions --- the one that minimizes the pressure gradient cost, which matches the philosophy of Gauss' principle. 
In the case of a circular cylinder, the solution that minimizes the pressure gradient is the one with zero circulation \cite{gonzalez2022variational}.

\begin{figure}[!]
\centering
\includegraphics[width=0.4\textwidth]{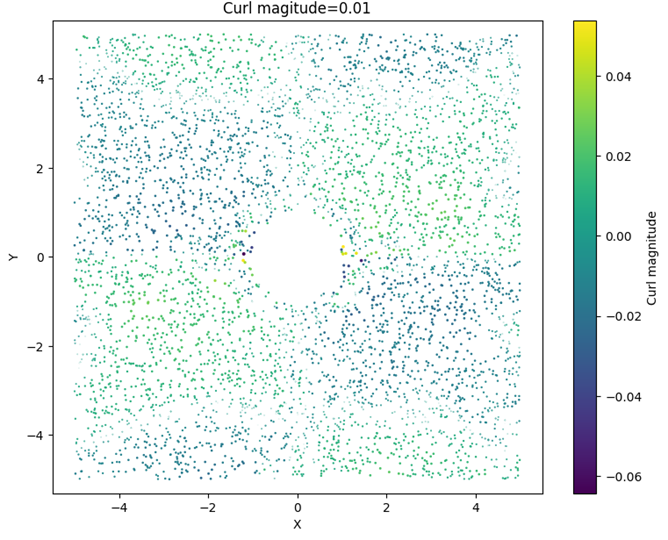}
\caption{\label{fig: Curl map}  Contours of the Curl of convective acceleration.}
\end{figure}

\begin{figure*}[!]
\centering
\includegraphics[width=0.9\textwidth]{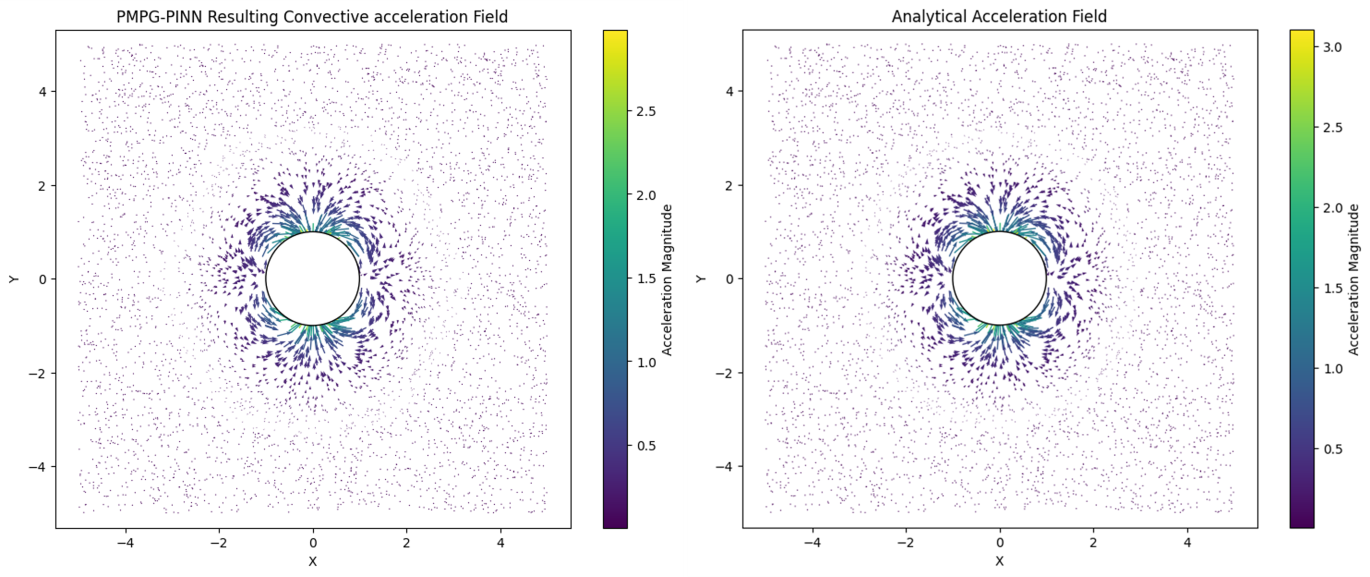}
\caption{\label{fig:acceleratation field} PMPG-PINN Resulting Convective Acceleration field vs Analytical solution Convective Acceleration field.}
\end{figure*}

\paragraph{The curl of the acceleration vector} provides a measure of satisfying the conservation of momentum at equilibrium.
In the case of a steady, inviscid flow, the curl of the convective acceleration should be zero. 
Our model was constrained to satisfy this equilibrium condition.
The resulting RMS of this quantity was found to be 0.01, indicating excellent agreement with the analytical solution, as shown in Figure \ref{fig: Curl map}.
Figure \ref{fig:acceleratation field} compares the convective acceleration fields between the resulting flow field and the analytical solution.

\paragraph{The Pressure Gradient} 
The incorporation of the Principle of Minimum Pressure Gradient (PMPG) into our approach required adding the total magnitude of the pressure gradient as an additional term in the cost function; that is the sum of the squared of the magnitude of the pressure gradient (the convective acceleration in the steady, inviscid case):
\begin{equation}
S = \sum \left| (\mathbf{u} \cdot \nabla) \mathbf{u} \right|^2 ,
\label{eq:sum_of_pressure_gradients}
\end{equation}
which is called "\textit{Appellian}". 
The normalized Appellian is computed as a physical metric and compared to the one computed for the analytical solution.



\section{\label{sec:Comparison}Comparison with different techniques}

To gain insight into the strengths and limitations of the proposed approach, we compare the performance of four different cost functions using: (a) our approach, which includes a weighted sum between the pressure gradient cost and the equilibrium loss function (represented by the curl of the convective acceleration), (b) equilibrium only, (c) the pressure gradient only, and (d) traditional PINNs.
\paragraph{PMPG and Equilibrium}  This is the proposed approach used to produce the resulting flow field presented above, which was described in details in Section \ref{sec:PMPG based PINNs}.
The corresponding cost function is presented in Equation \ref{eq:PMPG + curl}.

\paragraph{Equilibrium (Conservation of Momentum) only}  The second variant is imposing the conservation of momentum as well as satisfying the boundary conditions without incorporating the pressure gradient into the cost function given:
\begin{equation}
\label{eq:curl only}
\mathcal{L}(\mathbf{x};\boldsymbol{\theta}) = \lambda_1 \mathcal{L}_{\text{BC}}(\mathbf{x};\boldsymbol{\theta})+ \lambda_2 \mathcal{L}_{\text{Eqm}}(\mathbf{x};\boldsymbol{\theta}).
\end{equation}

\paragraph{PMPG only}  The third case is to minimize the pressure gradient while satisfying the boundary conditions without imposing conservation of momentum as a hard constraint on the curl of the convective acceleration. 
In this case, the cost function is given by :
\begin{eqnarray}
\label{eq:pmpg only}
\mathcal{L}(\mathbf{x};\boldsymbol{\theta}) = \lambda_1 \mathcal{L}_{\text{BC}}(\mathbf{x};\boldsymbol{\theta})+\lambda_3\mathcal{L}_{PMPG}(\mathbf{x};\boldsymbol{\theta}).
\end{eqnarray}
It should be emphasized that in the unsteady application of this approach, such a constraint would not be needed because the minimizing solution would naturally satisfy conservation of momentum as proved by Taha et al. \cite{taha2023}.
However, a similar fact in the steady case may not be true, as demonstrated in the ensuing comparative analysis.

\paragraph{Traditional PINNs}  The fourth case is added to compare the performance of the proposed approach to traditional PINNs where an additional network is constructed for the pressure field and conservation of momentum is imposed by minimizing the residuals of the equation (i.e, the difference between the pressure gradient and the convective acceleration in our case).
The same Neural Network structure, discussed in section \ref{sec:NN structure}, was used to model the pressure.
The boundary conditions are imposed in the same way as in the previous three cases.
As such, the cost function is given by:

\begin{eqnarray}
\label{eq:classical pinns}
\mathcal{L}(\mathbf{x};\boldsymbol{\theta}) = \lambda_1 \mathcal{L}_{\text{BC}}(\mathbf{x};\boldsymbol{\theta})+\lambda_4\mathcal{L}_{PINNs}(\mathbf{x};\boldsymbol{\theta}),
\end{eqnarray}
where the $\mathcal{L}_{PINNs}$ is given by:
\begin{eqnarray}
\label{eq: pinns}
\mathcal{L}_{PINNs} = \sum \left| (\mathbf{u} \cdot \nabla) \mathbf{u} + \frac{\nabla P}{\rho}  \right|^2.
\end{eqnarray}
The training of the four previous cases was monitored and recorded after each 25 epochs for 30000 epochs. 
The following is a detailed comparison in terms of different metrics.

\subsubsection{\label{subsec:Training time}Training time}
All cases were trained for 30000 epochs on the same processor performance. As observed in Table \ref{tab:training times}, the training time of the proposed formulation of \textit{PMPG and Equilibrium} was the fastest, with a slight advantage over both the \textit{Equilibrium only} and \textit{PMPG only} cases. 
Specifically, the training time of \textit{Traditional PINNs} was found to be approximately 43\% higher than that of the proposed approach, \textit{PMPG and Equilibrium}. 
This result highlights the efficacy of the proposed approach in reducing training time.
Although not handled here, it is expected that this superiority of the PMPG-PINN approach over the traditional PINNs will increase in scenarios involving complex geometries, larger number of points, more sophisticated Neural Network structure, and in three-dimensional and unsteady cases.
\begin{table}[ht]
\caption{\label{tab:training times}Training times comparison and RMS error of the resulting velocity fields for each case.}
\begin{ruledtabular}
\begin{tabular}{cccddd}
Case&Training time (seconds)& RMS Error\\
\hline
PMPG and Equilibrium & 1080.67 & 1.8\\
Equilibrium only & 1088.62 & 3.7\\
PMPG only & 1088.96 & 16.5\\
Traditional PINNs & 1546.05 & 4.8\\
\end{tabular}
\end{ruledtabular}
\end{table}
\subsubsection{\label{subsec:RMSE}Root Mean Square Error}
\begin{figure}[!]
\centering
\includegraphics[width=0.4\textwidth]{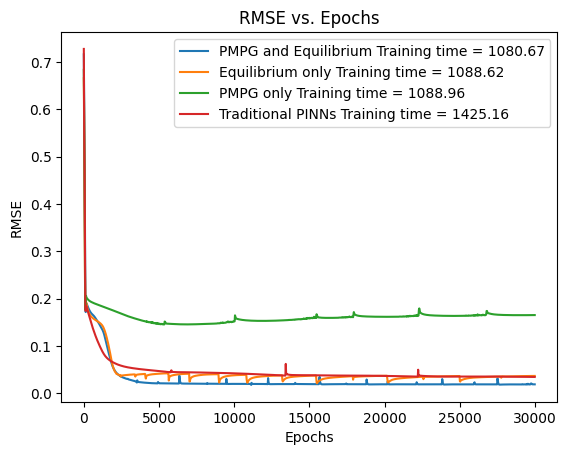}
\caption{\label{fig: RMSE compare}  A comparison of the RMS error of the velocity field among the four cases of study vs Epochs: \textit{PMPG and Equilibrium}, \textit{Equilibrium only, PMPG only} and \textit{Traditional PINNs}}
\end{figure}

\begin{figure}[!]
\centering
\includegraphics[width=0.45\textwidth]{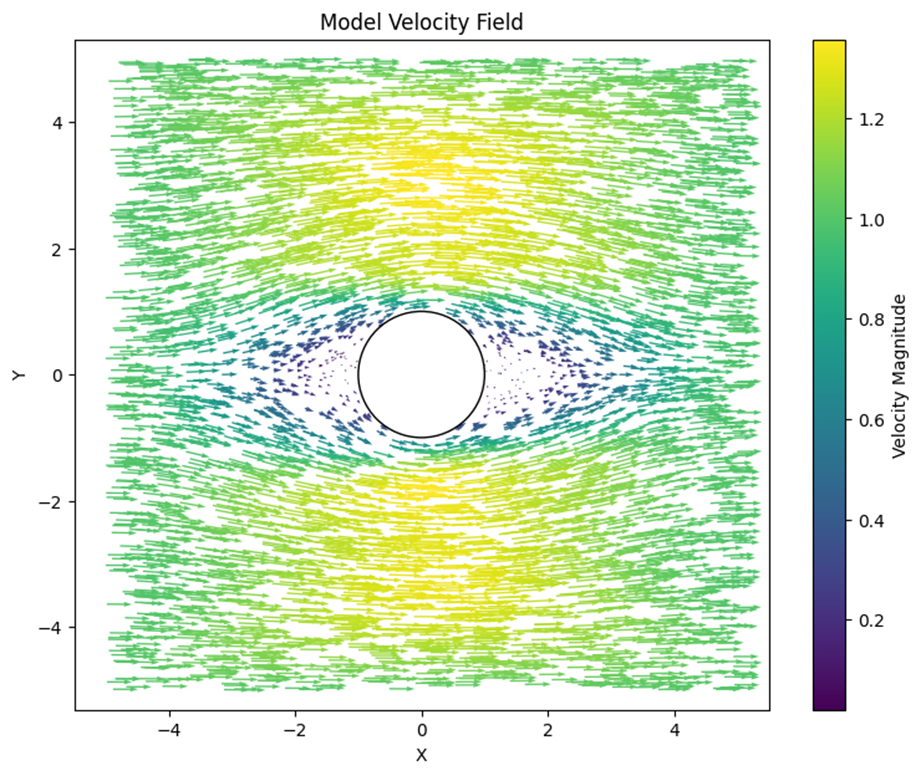}
\caption{\label{fig:PMPG velocity profile} Velocity Field of the non-physical case of \textit{PMPG only}.}
\end{figure}

Figure \ref{fig: RMSE compare} depicts the RMS error of all cases.
The proposed training approach (\textit{PMPG and Equilibrium}) [Section \ref{sec:error analysis}] achieves a velocity field error of 1.8\%, the lowest among all cases.
The RMS error of \textit{Equilibrium only} was 3.7\%, almost double that of the proposed approach while the \textit{Traditional PINNs} approach results in an RMS error of 4.8\%.
Both the \textit{Equilibrium only} and \textit{Traditional PINNs} approaches yielded higher errors compared to the \textit{PMPG and Equilibrium} approach.
On the other hand, the \textit{PMPG only} case resulted in a nonphysical flow field as shown in Figure \ref{fig:PMPG velocity profile}, with an RMS error of 16.5\%. 
This discrepancy is attributed to the fact that the solution that the \textit{Steady} flow minimizes (the pressure gradient cost) may not necessarily satisfy conservation of momentum, in contrast to the unsteady case where the pressure gradient cost is minimized at every time instant of time.
In this case, the resulting flow field is guaranteed to naturally satisfy the conservation of momentum as ensured by Theorem 2 in Taha et al. work \cite{taha2023}.
Therefore, in the steady case, equilibrium must be explicitly enforced.
Otherwise, the resulting flow field may be non-physical.
Section \ref{sec:Physics-Driven regularization} will discuss this point from a different perspective, related to the absence of a regularization term. 

\subsubsection{\label{subsec:Physics checks1}Physical Measures: Vorticity and Equilibrium}

\begin{figure*}[!]
\centering
\includegraphics[width=0.9\textwidth]{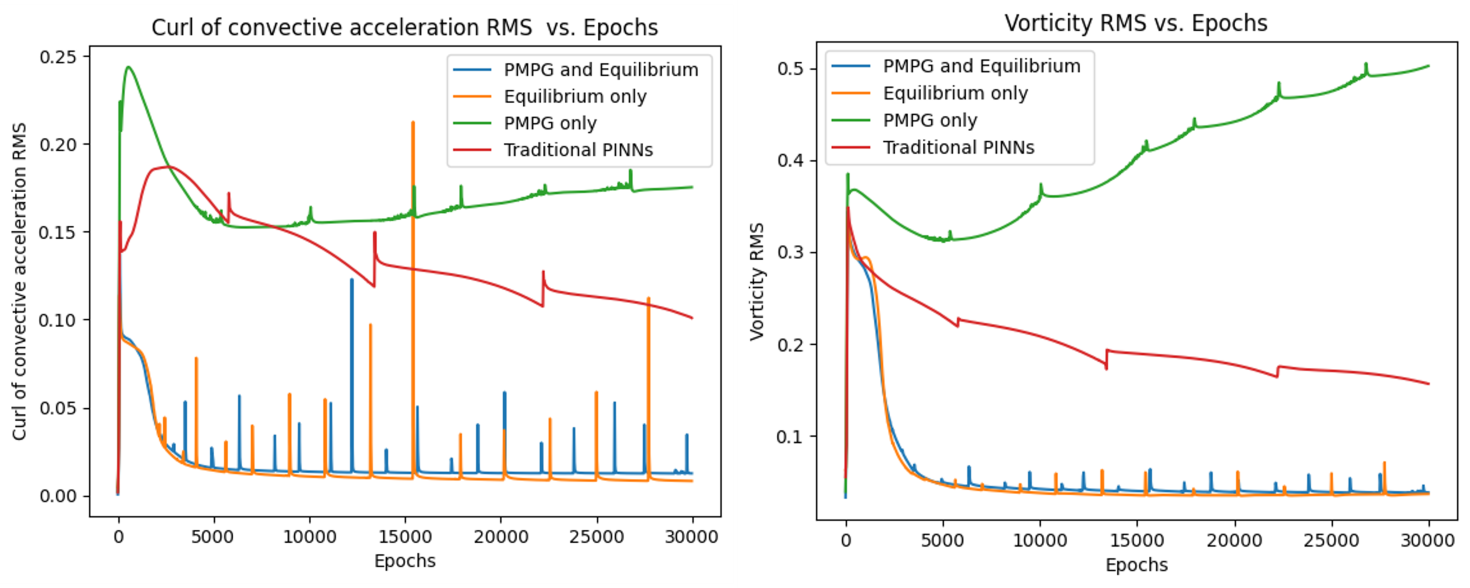}
\caption{\label{fig:Vorticity and curl compare} Vorticity and curl of convective acceleration vs Epochs for the four cases of study.}
\end{figure*}

Figure \ref{fig:Vorticity and curl compare} compares the RMS of both the vorticity and the curl of the convective acceleration for all cases.
As mentioned earlier in Section \ref{subsec:RMSE}, the \textit{PMPG only} does not guarantee that the steady flow field satisfies the equilibrium equation.
Hence, the corresponding RMS of the curl of the convective acceleration is significantly higher than all other cases, as shown in Figure \ref{fig:Vorticity and curl compare}, because of the violation of conservation of momentum.
It also results in largely vortical flow fields, which further reinforces our understanding that minimizing the pressure gradient in a steady problem without imposing an equilibrium constraint is insufficient.
The values of the curl of acceleration and vorticity magnitudes for both the proposed approach, \textit{PMPG and Equilibrium} and \textit{Equilibrium only} are acceptable and considerably better than the results from the \textit{Traditional PINNs} approach.
It should be noted that the \textit{Traditional PINNs} training exhibits a slow convergence trend which may require more iterations to reach acceptable levels.

\subsubsection{\label{subsec:Pressure Gradient}Pressure Gradient}
\begin{figure}[!]
\centering
\includegraphics[width=0.4\textwidth]{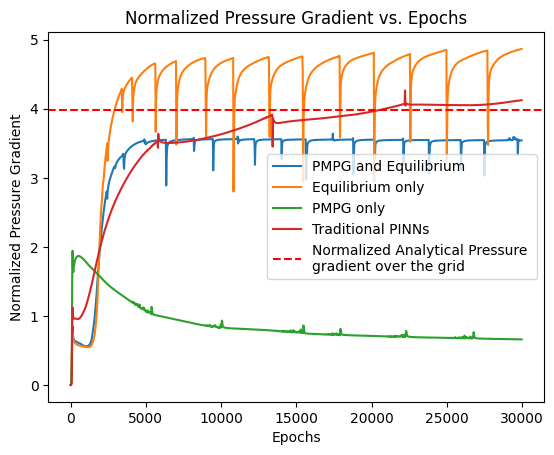}
\caption{\label{fig: Pressure Gradient compare} Normalized Pressure Gradient of the four cases of study vs analytical solution over the same domain.}
\end{figure}

As discussed earlier, the minimization of the pressure gradient without imposing an equilibrium constraint (in steady cases) is insufficient.
Similarly, imposing the equilibrium constraint alone is also not enough, as demonstrated by both the \textit{Equilibrium only} and \textit{Traditional PINNs} cases, particularly for Euler flows where there may exist multiple equilibria, but only one is physical --- the one that minimizes the pressure gradient.
Figure \ref{fig: Pressure Gradient compare} presents a comparison of the Normalized Appellian variation during iterations among the four cases of study.
It shows that the \textit{Equilibrium only} case asymptotes to a pressure gradient value higher than that of the analytical solution. 
The \textit{Traditional PINNs} approach results in a higher pressure gradient too.
As may be expected, the proposed approach \textit{PMPG and Equilibrium}, converges to a solution with a smaller Appellian than the \textit{Equilibrium only} and \textit{Traditional PINNs} approaches.
Also, the \textit{PMPG only} approach clearly converges to a solution with the least Appellian, but it is non-physical as it violates conservation of momentum.
It is very well known in optimization that relaxing constraints (equilibrium in this case) yields a smaller optimal cost \cite{arora2015optimization}.
This behavior implies that the equilibrium constraint is an \textit{active} constraint that drives the optimization process, therefore, it must be taken into consideration.
The gap between the theoretical pressure gradient and that of the \textit{PMPG and Equilibrium} is attributed to the slight error in the velocity field.
With a more suitable neural network structure, an adjusted domain size, and increased collocation point density, we believe this gap can be further reduced.

\subsubsection{\label{subsec:Convergence}Convergence}
\begin{figure}[!]
\centering
\includegraphics[width=0.4\textwidth]{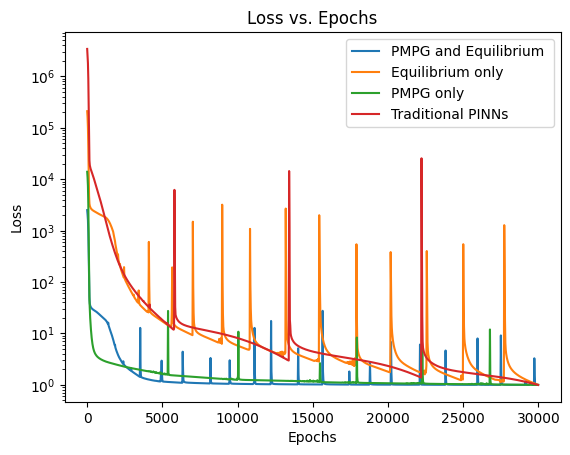}
\caption{\label{fig: Total loss}  Total loss (normalized by the final value) for the four cases of study as an indication of convergence rate.}
\end{figure}

While we anticipate different values for the total loss for each case due to the application of various constraints, which are enforced using the penalty method, it is evident from Figure \ref{fig: Total loss} and other figures that the \textit{PMPG and Equilibrium} approach converges the fastest.

In summary, the proposed approach of \textit{PMPG and Equilibrium} outperforms all of the other three approaches in terms of the RMS error of the velocity field, training time, convergence rate, and compliance with physics.
While the studied numerical example features a classic, simple geometry, which may preclude the generalization of the above conclusions, these advantages are likely to persist with more complex geometries and flow fields.
We specifically chose this simple example to gain insight into the strengths of the PMPG-PNN approach in comparison to the other traditional PINN approaches without blurring the picture with the details of a complex flow field.
However, the ability of the proposed approach to tackle more complex fluid problems remains to be assessed in future work.
For these cases, several other factors could further enhance the accuracy of the results, such as increasing the density of collocation points, adjusting the domain size, selecting a more suitable optimizer and activation function, and employing more complex Neural Network structures.

\section{\label{sec:Numerics}Numerical analysis}

In this section, we examine the sensitivity of the resulting flow field from the proposed approach of the \textit{PMPG and Equilibrium} to variations in the number of neurons in the neural network architecture. 
Since the model is data-free, there is no validation set as typically used in conventional neural network training. 
Instead, we rely on theoretical data to validate the model and optimize the number of neurons. 
Additionally, we analyze the sensitivity of the coefficient $\lambda_3$, which controls the weight assigned to the PMPG loss term $\mathcal{L}_{PMPG}(\mathbf{x}; \boldsymbol{\theta})$.

\subsubsection{\label{subsec:neoruns_2}Effect of Number of neurons}

To optimize the number of neurons for $\psi$ that can accurately approximate the analytical solution, we use the same neural network structure shown in Figure \ref{fig:NN structure}. 
The network takes the two spatial coordinates, $x$ and $y$, as inputs and outputs a single value for $\psi$. 
The cost function used to evaluate the network's performance is the RMS error between the predicted and analytical velocity fields. 
We conduct tests using different numbers of neurons in the hidden layer to determine the optimal configuration.

\begin{figure}[!]
\centering
\includegraphics[width=0.4\textwidth]{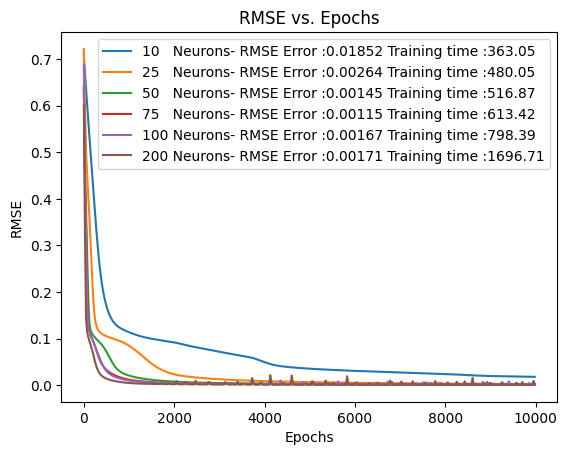}
\caption{\label{fig: number of neurons1}  RMS error of the Resulting velocity field vs epochs for different Numbers of Neurons.}
\end{figure}

\begin{figure}[!]
\centering
\includegraphics[width=0.4\textwidth]{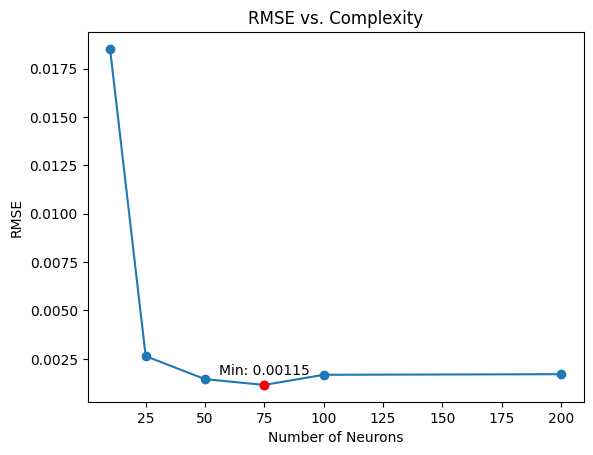}
\caption{\label{fig: number of neurons2}  RMS error of the Resulting Velocity field vs Number of Neurons for 10000 epochs.}
\end{figure}

The number of neurons tested were 10, 25, 50, 75, 100, and 200. 
Each configuration was trained for 10,000 epochs, as shown in Figure \ref{fig: number of neurons1}.
As expected, training time increased with the complexity of the neural network as the number of neurons grew. 
The lowest RMS error was observed with 75 neurons, achieving an RMS error of 0.115\%, which provided a slight advantage over the configurations with 50 and 100 neurons, as illustrated in Figure \ref{fig: number of neurons2}.

\subsubsection{\label{subsec:neoruns}Effect of the Weight of the PMPG Loss}

\begin{figure}[!]
\centering
\includegraphics[width=0.4\textwidth]{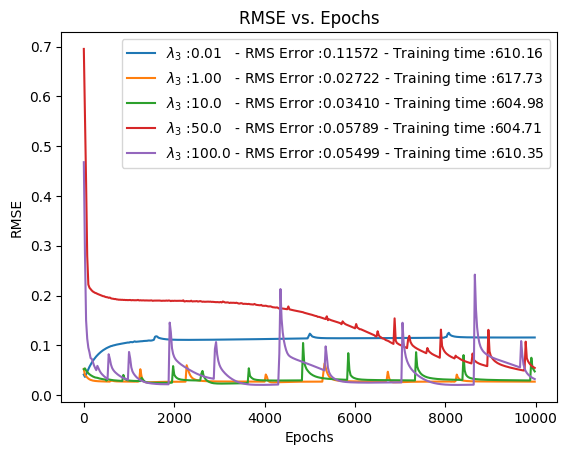}
\caption{\label{fig: coeeficient1}  RMS error of the Resulting velocity field vs epochs for different $\lambda_3$ values using the 75-Neurons Neural Network.}
\end{figure}

\begin{figure}[!]
\centering
\includegraphics[width=0.4\textwidth]{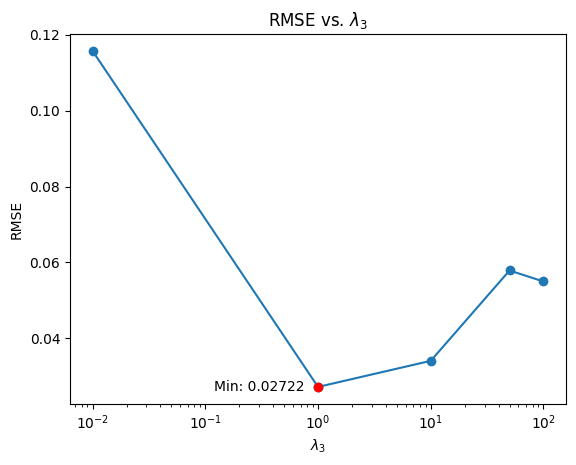}
\caption{\label{fig: coeeficient2}  RMS error for the Resulting velocity field vs $\lambda_3$ for 10000 epochs using the 75-Neurons Neural Network.}
\end{figure}

Another tuning parameter that is worthy of investigation is the $\lambda_3$ in Equation \ref{eq:PMPG + curl} i.e., the weighting coefficient of the PMPG Loss in the total loss function (while fixing the weights of all other losses at unity values).
Different values for the weight $\lambda_3$ were tested: 0.01, 1, 10, 50 and 100, and the model was trained for 10000 epochs using the 75-neuron Neural Network structure. 
Figure \ref{fig: coeeficient1} shows the variation of RMS error with epochs at each value of $\lambda_3$, while Figure \ref{fig: coeeficient2} shows the variation of the converged solutions after 10000 epochs with the weight $\lambda_3$.
The latter figure implies that the least RMS error of the velocity field is achieved when $\lambda_3$ =1 with an RMS error of 2.7\%.

\subsubsection{\label{subsec:Activation function}Activation function}

Due to the smooth nature of the expected solution of the problem selected for this study, replacing the activation function (the hyperbolic tangent (Tanh) function) with other activation functions such as the Rectified Linear Unit (\textit{ReLU})\cite{petersen2018optimal} or its variants such as the Leaky Rectified Linear Unit (\textit{LReLU}) and the Exponential Linear Unit (\textit{ELU}) was not inspected. 
The use of these discontinuous activation functions can allow for a wider space of solutions that can capture more complex physical phenomena, such as flow separation or generation of sheets of discontinuity (i.e., vortex sheets) in an unsteady problem (e.g., starting flow over an airfoil). 

\section{\label{sec:Physics-Driven regularization}Physics-Driven regularization}

Regularization is a fundamental technique in machine learning used to mitigate overfitting, enhance model generalization, and improve predictive accuracy, especially in problems characterized by non-uniqueness of the solution \cite{heaton2018ian}. 
In conventional data-driven machine learning models, regularization methods (such as L1 and L2 penalties, dropout, and early stopping) are employed to control the complexity of the model by penalizing large weights, thereby preventing the model from following the noise in the training data and thus enhancing the accuracy of generalization.
In other words, regularization reduces complexity of the Neural Network by smoothing the target function to make it follow the average trend rather than the sparse data.
This smoothing is usually performed by incorporating the derivatives of the weights of the Neural Network into the cost function via the so-called "Regularization term".
Validation techniques, including cross-validation, are crucial in this process as they allow for the assessment of model performance on unseen data, ensuring that the model generalizes well beyond the training dataset.

In the context of our "Data-Free" physics-driven Neural Networks, traditional validation techniques are not applicable due to the absence of empirical data. Our model is designed to solve physics problems solely based on the governing equations of the physical system, without relying on external datasets for training or validation. 
As a result, the standard approach of splitting the available data set into training, validation, and test sets is not feasible for the proposed approach.

The challenge of overfitting remains a dilemma for the PMPG-PINN model, and the absence of data necessitates alternative regularization methods. 
However, enforcing the equilibrium condition as a physics constraint to the optimization (using the curl of the convective acceleration as a penalty constraint) can play the role of a regularization penalty. 
In this physics-driven regularization, the regularizing penalty comes from a physical constraint, thereby guiding the model to generalize appropriately within the defined physical framework without overfitting.

There is a notable parallel between standard regularization techniques and the incorporation of the Equilibrium as a constraint within the proposed framework of this work.
This resemblance suggests that the Equilibrium penalty in Equation \ref{eq:PMPG + curl} fulfills dual roles—both physical and numerical—throughout the training process, ultimately guiding the model smoothly towards the physical solution.

To highlight this connection, the standard formulation of regularization in Neural Network training problems is as follows:
\begin{equation}
\begin{aligned}
\min_{\theta} \quad \mathcal{L} (\mathbf{x};\boldsymbol{\theta}) + \lambda \Omega(\theta),
\end{aligned}
\end{equation}
where \(\lambda\) denotes the regularization coefficient, \(\mathcal{L} (\mathbf{x};\boldsymbol{\theta})\) is the objective cost function, and \(\Omega(\theta)\) is a functional involving the weights \(\theta\) and their derivatives. 
Increasing \(\lambda\) imposes greater smoothness on the model by penalizing deviations.
This concept is derived from a constrained minimization problem, where smoothness requirements are enforced through the penalty method.

Input gradient regularization \cite{finlay2021scaleable} is one of the regularization techniques, where the regularization term \(\Omega(\cdot)\) is a function of the gradient of the loss function with respect to the inputs, \(\nabla_{\mathbf{x}} \mathcal{L} (\mathbf{x};\boldsymbol{\theta})\). 
Here, \(\mathbf{x}\) represents the neural network inputs. 
This technique mirrors the equilibrium term \(\mathcal{L}_{\text{Eqm}}(\mathbf{x};\boldsymbol{\theta})\), weighted by \(\lambda_2\) in Equation \ref{eq:PMPG + curl}, as both involve derivatives of a formulated cost function and both are imposed by the penalty method.
Specifically, the curl of convective acceleration can be analogous to the derivatives of the cost function, akin to how regularization terms achieves smoothness in standard regularization techniques.

\section{\label{sec:Conclusions}Conclusions}

In this paper, we propose a new approach for simulating fluid dynamic problems by integrating Physics-Informed Neural Networks (PINNs) and the Principle of Minimum Pressure Gradient (PMPG).
The approach is a "Data-Free" technique and therefore does not require extensive data for training
The proposed approach is demonstrated using a simple yet illustrative example: the steady, ideal flow around a two-dimensional cylinder. 
This simple example was particularly selected to gain insight into the strengths and limitations of the proposed approach in comparison to standard PINN techniques, without cluttering the picture with complex flow details.
Utilizing PMPG represents a significant advancement in streamlining the training process of PINNs, resulting in notable reductions in computational costs by eliminating the need for a separate pressure model. 
By minimizing a physically meaningful quantity, guided by the PMPG, as opposed to employing traditional least squares residuals, our meshless technique significantly reduces both the time and effort typically required for grid generation.

Through the analysis shown in this research, the PMPG-PINN outperforms the traditional PINNs in terms of (i) convergence rate, (ii) training time, (iii) accuracy, and (iv) matching with physics.
The analysis conducted in this research also shows that relying solely on minimizing the pressure gradient in steady problems may be inadequate, as the converged solution may not necessarily satisfy equilibrium equations.

Furthermore, the proposed approach offers a promising avenue for filtering noisy data fields, such as those obtained from experimental Particle Image Velocimetry (PIV) or echo PIV techniques. Augmenting the cost function with imported data facilitates the incorporation of filtering mechanisms to remove outliers or noise through physical corrections, mirroring the methodologies employed in this study.

The versatility of the proposed methodology extends beyond the presented example, readily adaptable to more intricate scenarios encompassing three-dimensional, unsteady, viscous flows involving non-Newtonian fluids subjected to arbitrary external forces.
Future investigations should assess the computational efficiency of the PMPG-PINN technique in handling more complex flow fields.

\nocite{*}

\begin{thebibliography}{70}%
\makeatletter
\providecommand \@ifxundefined [1]{%
 \@ifx{#1\undefined}
}%
\providecommand \@ifnum [1]{%
 \ifnum #1\expandafter \@firstoftwo
 \else \expandafter \@secondoftwo
 \fi
}%
\providecommand \@ifx [1]{%
 \ifx #1\expandafter \@firstoftwo
 \else \expandafter \@secondoftwo
 \fi
}%
\providecommand \natexlab [1]{#1}%
\providecommand \enquote  [1]{``#1''}%
\providecommand \bibnamefont  [1]{#1}%
\providecommand \bibfnamefont [1]{#1}%
\providecommand \citenamefont [1]{#1}%
\providecommand \href@noop [0]{\@secondoftwo}%
\providecommand \href [0]{\begingroup \@sanitize@url \@href}%
\providecommand \@href[1]{\@@startlink{#1}\@@href}%
\providecommand \@@href[1]{\endgroup#1\@@endlink}%
\providecommand \@sanitize@url [0]{\catcode `\\12\catcode `\$12\catcode `\&12\catcode `\#12\catcode `\^12\catcode `\_12\catcode `\%12\relax}%
\providecommand \@@startlink[1]{}%
\providecommand \@@endlink[0]{}%
\providecommand \url  [0]{\begingroup\@sanitize@url \@url }%
\providecommand \@url [1]{\endgroup\@href {#1}{\urlprefix }}%
\providecommand \urlprefix  [0]{URL }%
\providecommand \Eprint [0]{\href }%
\providecommand \doibase [0]{http://dx.doi.org/}%
\providecommand \selectlanguage [0]{\@gobble}%
\providecommand \bibinfo  [0]{\@secondoftwo}%
\providecommand \bibfield  [0]{\@secondoftwo}%
\providecommand \translation [1]{[#1]}%
\providecommand \BibitemOpen [0]{}%
\providecommand \bibitemStop [0]{}%
\providecommand \bibitemNoStop [0]{.\EOS\space}%
\providecommand \EOS [0]{\spacefactor3000\relax}%
\providecommand \BibitemShut  [1]{\csname bibitem#1\endcsname}%
\let\auto@bib@innerbib\@empty
\bibitem [{\citenamefont {Raissi}, \citenamefont {Perdikaris},\ and\ \citenamefont {Karniadakis}(2019)}]{raissi_physics-informed_2019}%
  \BibitemOpen
  \bibfield  {author} {\bibinfo {author} {\bibfnamefont {M.}~\bibnamefont {Raissi}}, \bibinfo {author} {\bibfnamefont {P.}~\bibnamefont {Perdikaris}}, \ and\ \bibinfo {author} {\bibfnamefont {G.}~\bibnamefont {Karniadakis}},\ }\bibfield  {title} {{\selectlanguage {en}\enquote {\bibinfo {title} {Physics-informed neural networks: {A} deep learning framework for solving forward and inverse problems involving nonlinear partial differential equations},}\ }}\href {\doibase 10.1016/j.jcp.2018.10.045} {\bibfield  {journal} {\bibinfo  {journal} {Journal of Computational Physics}\ }\textbf {\bibinfo {volume} {378}},\ \bibinfo {pages} {686--707} (\bibinfo {year} {2019})}\BibitemShut {NoStop}%
\bibitem [{\citenamefont {Seliger}\ and\ \citenamefont {Whitham}(1968)}]{seliger1968}%
  \BibitemOpen
  \bibfield  {author} {\bibinfo {author} {\bibfnamefont {R.~L.}\ \bibnamefont {Seliger}}\ and\ \bibinfo {author} {\bibfnamefont {G.~B.}\ \bibnamefont {Whitham}},\ }\bibfield  {title} {\enquote {\bibinfo {title} {Variational principles in continuum mechanics},}\ }\href {\doibase 10.1098/rspa.1968.0031} {\bibfield  {journal} {\bibinfo  {journal} {Proceedings of the Royal Society of London A: Mathematical, Physical and Engineering Sciences}\ }\textbf {\bibinfo {volume} {305}},\ \bibinfo {pages} {1--25} (\bibinfo {year} {1968})}\BibitemShut {NoStop}%
\bibitem [{\citenamefont {Taha}\ and\ \citenamefont {Gonzalez}(2021)}]{taha2022does}%
  \BibitemOpen
  \bibfield  {author} {\bibinfo {author} {\bibfnamefont {H.~E.}\ \bibnamefont {Taha}}\ and\ \bibinfo {author} {\bibfnamefont {C.}~\bibnamefont {Gonzalez}},\ }\bibfield  {title} {\enquote {\bibinfo {title} {What does nature minimize in every incompressible flow?}}\ }\href@noop {} {\bibfield  {journal} {\bibinfo  {journal} {arXiv preprint arXiv:2112.12261}\ } (\bibinfo {year} {2021})}\BibitemShut {NoStop}%
\bibitem [{\citenamefont {Taha}, \citenamefont {Gonzalez},\ and\ \citenamefont {Shorbagy}(2023)}]{taha2023}%
  \BibitemOpen
  \bibfield  {author} {\bibinfo {author} {\bibfnamefont {H.}~\bibnamefont {Taha}}, \bibinfo {author} {\bibfnamefont {C.}~\bibnamefont {Gonzalez}}, \ and\ \bibinfo {author} {\bibfnamefont {M.}~\bibnamefont {Shorbagy}},\ }\bibfield  {title} {\enquote {\bibinfo {title} {A minimization principle for incompressible fluid mechanics},}\ }\href@noop {} {\bibfield  {journal} {\bibinfo  {journal} {Physics of Fluids}\ }\textbf {\bibinfo {volume} {35}} (\bibinfo {year} {2023})}\BibitemShut {NoStop}%
\bibitem [{\citenamefont {Atallah}, \citenamefont {Elmaradny},\ and\ \citenamefont {Taha}(2024)}]{atallah2024novel}%
  \BibitemOpen
  \bibfield  {author} {\bibinfo {author} {\bibfnamefont {A.}~\bibnamefont {Atallah}}, \bibinfo {author} {\bibfnamefont {A.}~\bibnamefont {Elmaradny}}, \ and\ \bibinfo {author} {\bibfnamefont {H.~E.}\ \bibnamefont {Taha}},\ }\bibfield  {title} {\enquote {\bibinfo {title} {A novel approach for data-free, physics-informed neural networks in fluid mechanics using the principle of minimum pressure gradient},}\ }in\ \href@noop {} {\emph {\bibinfo {booktitle} {AIAA SCITECH 2024 Forum}}}\ (\bibinfo {year} {2024})\ p.\ \bibinfo {pages} {2742}\BibitemShut {NoStop}%
\bibitem [{\citenamefont {Alhussein}\ and\ \citenamefont {Daqaq}(2024)}]{alhussein2024principle}%
  \BibitemOpen
  \bibfield  {author} {\bibinfo {author} {\bibfnamefont {H.}~\bibnamefont {Alhussein}}\ and\ \bibinfo {author} {\bibfnamefont {M.}~\bibnamefont {Daqaq}},\ }\bibfield  {title} {\enquote {\bibinfo {title} {{The principle of minimum pressure gradient: An alternative basis for physics-informed learning of incompressible fluid mechanics}},}\ }\href {\doibase 10.1063/5.0197860} {\bibfield  {journal} {\bibinfo  {journal} {AIP Advances}\ }\textbf {\bibinfo {volume} {14}},\ \bibinfo {pages} {045112} (\bibinfo {year} {2024})},\ \Eprint {http://arxiv.org/abs/https://pubs.aip.org/aip/adv/article-pdf/doi/10.1063/5.0197860/19870776/045112\_1\_5.0197860.pdf} {https://pubs.aip.org/aip/adv/article-pdf/doi/10.1063/5.0197860/19870776/045112\_1\_5.0197860.pdf} \BibitemShut {NoStop}%
\bibitem [{\citenamefont {Bretherton}(1970)}]{bretherton1970}%
  \BibitemOpen
  \bibfield  {author} {\bibinfo {author} {\bibfnamefont {F.~P.}\ \bibnamefont {Bretherton}},\ }\bibfield  {title} {\enquote {\bibinfo {title} {A note on hamilton’s principle for perfect fluids},}\ }\href {\doibase 10.1017/S0022112070001275} {\bibfield  {journal} {\bibinfo  {journal} {Journal of Fluid Mechanics}\ }\textbf {\bibinfo {volume} {44}},\ \bibinfo {pages} {19} (\bibinfo {year} {1970})}\BibitemShut {NoStop}%
\bibitem [{\citenamefont {Salmon}(1988)}]{salmon1988}%
  \BibitemOpen
  \bibfield  {author} {\bibinfo {author} {\bibfnamefont {R.}~\bibnamefont {Salmon}},\ }\bibfield  {title} {\enquote {\bibinfo {title} {Hamiltonian fluid mechanics},}\ }\href {\doibase 10.1146/annurev.fl.20.010188.001301} {\bibfield  {journal} {\bibinfo  {journal} {Annual Review of Fluid Mechanics}\ }\textbf {\bibinfo {volume} {20}},\ \bibinfo {pages} {225} (\bibinfo {year} {1988})}\BibitemShut {NoStop}%
\bibitem [{\citenamefont {Morrison}(1998)}]{morrison1998}%
  \BibitemOpen
  \bibfield  {author} {\bibinfo {author} {\bibfnamefont {P.~J.}\ \bibnamefont {Morrison}},\ }\bibfield  {title} {\enquote {\bibinfo {title} {Hamiltonian description of the ideal fluid},}\ }\href {\doibase 10.1103/RevModPhys.70.467} {\bibfield  {journal} {\bibinfo  {journal} {Reviews of Modern Physics}\ }\textbf {\bibinfo {volume} {70}},\ \bibinfo {pages} {467} (\bibinfo {year} {1998})}\BibitemShut {NoStop}%
\bibitem [{\citenamefont {Fukagawa}\ and\ \citenamefont {Fujitani}(2012)}]{fukagawa2012variational}%
  \BibitemOpen
  \bibfield  {author} {\bibinfo {author} {\bibfnamefont {H.}~\bibnamefont {Fukagawa}}\ and\ \bibinfo {author} {\bibfnamefont {Y.}~\bibnamefont {Fujitani}},\ }\bibfield  {title} {\enquote {\bibinfo {title} {A variational principle for dissipative fluid dynamics},}\ }\href@noop {} {\bibfield  {journal} {\bibinfo  {journal} {Progress of Theoretical Physics}\ }\textbf {\bibinfo {volume} {127}},\ \bibinfo {pages} {921--935} (\bibinfo {year} {2012})}\BibitemShut {NoStop}%
\bibitem [{\citenamefont {Galley}, \citenamefont {Tsang},\ and\ \citenamefont {Stein}(2014)}]{galley2014principle}%
  \BibitemOpen
  \bibfield  {author} {\bibinfo {author} {\bibfnamefont {C.~R.}\ \bibnamefont {Galley}}, \bibinfo {author} {\bibfnamefont {D.}~\bibnamefont {Tsang}}, \ and\ \bibinfo {author} {\bibfnamefont {L.~C.}\ \bibnamefont {Stein}},\ }\bibfield  {title} {\enquote {\bibinfo {title} {The principle of stationary nonconservative action for classical mechanics and field theories},}\ }\href@noop {} {\bibfield  {journal} {\bibinfo  {journal} {arXiv preprint arXiv:1412.3082}\ } (\bibinfo {year} {2014})}\BibitemShut {NoStop}%
\bibitem [{\citenamefont {Gay-Balmaz}\ and\ \citenamefont {Yoshimura}(2018)}]{gay2018lagrangian}%
  \BibitemOpen
  \bibfield  {author} {\bibinfo {author} {\bibfnamefont {F.}~\bibnamefont {Gay-Balmaz}}\ and\ \bibinfo {author} {\bibfnamefont {H.}~\bibnamefont {Yoshimura}},\ }\bibfield  {title} {\enquote {\bibinfo {title} {A lagrangian variational formulation for nonequilibrium thermodynamics},}\ }\href@noop {} {\bibfield  {journal} {\bibinfo  {journal} {IFAC-PapersOnLine}\ }\textbf {\bibinfo {volume} {51}},\ \bibinfo {pages} {25--30} (\bibinfo {year} {2018})}\BibitemShut {NoStop}%
\bibitem [{\citenamefont {Sanders}\ \emph {et~al.}(2024)\citenamefont {Sanders}, \citenamefont {DeVoria}, \citenamefont {Washuta}, \citenamefont {Elamin}, \citenamefont {Skenes},\ and\ \citenamefont {Berlinghieri}}]{sanders2024canonical}%
  \BibitemOpen
  \bibfield  {author} {\bibinfo {author} {\bibfnamefont {J.~W.}\ \bibnamefont {Sanders}}, \bibinfo {author} {\bibfnamefont {A.~C.}\ \bibnamefont {DeVoria}}, \bibinfo {author} {\bibfnamefont {N.~J.}\ \bibnamefont {Washuta}}, \bibinfo {author} {\bibfnamefont {G.~A.}\ \bibnamefont {Elamin}}, \bibinfo {author} {\bibfnamefont {K.~L.}\ \bibnamefont {Skenes}}, \ and\ \bibinfo {author} {\bibfnamefont {J.~C.}\ \bibnamefont {Berlinghieri}},\ }\bibfield  {title} {\enquote {\bibinfo {title} {A canonical hamiltonian formulation of the navier--stokes problem},}\ }\href@noop {} {\bibfield  {journal} {\bibinfo  {journal} {Journal of Fluid Mechanics}\ }\textbf {\bibinfo {volume} {984}},\ \bibinfo {pages} {A27} (\bibinfo {year} {2024})}\BibitemShut {NoStop}%
\bibitem [{\citenamefont {Papastavridis}(2014)}]{papastavridis2014}%
  \BibitemOpen
  \bibfield  {author} {\bibinfo {author} {\bibfnamefont {J.}~\bibnamefont {Papastavridis}},\ }\href@noop {} {\emph {\bibinfo {title} {Analytical Mechanics: A Comprehensive Treatise on the Dynamics of Constrained Systems}}}\ (\bibinfo  {publisher} {World Scientific Publishing Company},\ \bibinfo {year} {2014})\ \bibinfo {note} {reprint edition}\BibitemShut {NoStop}%
\bibitem [{\citenamefont {Taha}\ and\ \citenamefont {Gonzalez}(2023{\natexlab{a}})}]{taha2023variational}%
  \BibitemOpen
  \bibfield  {author} {\bibinfo {author} {\bibfnamefont {H.~E.}\ \bibnamefont {Taha}}\ and\ \bibinfo {author} {\bibfnamefont {C.}~\bibnamefont {Gonzalez}},\ }\bibfield  {title} {\enquote {\bibinfo {title} {A variational principle for navier-stokes equations},}\ }in\ \href@noop {} {\emph {\bibinfo {booktitle} {AIAA SCITECH 2023 Forum}}}\ (\bibinfo {year} {2023})\ p.\ \bibinfo {pages} {1432}\BibitemShut {NoStop}%
\bibitem [{\citenamefont {Gonzalez}\ and\ \citenamefont {Taha}(2022)}]{gonzalez2022variational}%
  \BibitemOpen
  \bibfield  {author} {\bibinfo {author} {\bibfnamefont {C.}~\bibnamefont {Gonzalez}}\ and\ \bibinfo {author} {\bibfnamefont {H.~E.}\ \bibnamefont {Taha}},\ }\bibfield  {title} {\enquote {\bibinfo {title} {A variational theory of lift},}\ }\href@noop {} {\bibfield  {journal} {\bibinfo  {journal} {Journal of Fluid Mechanics}\ }\textbf {\bibinfo {volume} {941}},\ \bibinfo {pages} {A58} (\bibinfo {year} {2022})}\BibitemShut {NoStop}%
\bibitem [{\citenamefont {Taha}\ and\ \citenamefont {Gonzalez}(2022)}]{taha2022flow}%
  \BibitemOpen
  \bibfield  {author} {\bibinfo {author} {\bibfnamefont {H.~E.}\ \bibnamefont {Taha}}\ and\ \bibinfo {author} {\bibfnamefont {C.}~\bibnamefont {Gonzalez}},\ }\bibfield  {title} {\enquote {\bibinfo {title} {The flow over a flat plate: Did kutta get it right?}}\ }in\ \href@noop {} {\emph {\bibinfo {booktitle} {AIAA SCITECH 2022 Forum}}}\ (\bibinfo {year} {2022})\ p.\ \bibinfo {pages} {1665}\BibitemShut {NoStop}%
\bibitem [{\citenamefont {Taha}\ and\ \citenamefont {Gonzalez}(2023{\natexlab{b}})}]{taha2023refining}%
  \BibitemOpen
  \bibfield  {author} {\bibinfo {author} {\bibfnamefont {H.~E.}\ \bibnamefont {Taha}}\ and\ \bibinfo {author} {\bibfnamefont {C.}~\bibnamefont {Gonzalez}},\ }\bibfield  {title} {\enquote {\bibinfo {title} {Refining kutta’s flow over a flat plate: Necessary conditions for lift},}\ }\href@noop {} {\bibfield  {journal} {\bibinfo  {journal} {AIAA Journal}\ }\textbf {\bibinfo {volume} {61}},\ \bibinfo {pages} {2060--2068} (\bibinfo {year} {2023}{\natexlab{b}})}\BibitemShut {NoStop}%
\bibitem [{\citenamefont {Shorbagy}\ and\ \citenamefont {Taha}(2024)}]{shorbagy2024magnus}%
  \BibitemOpen
  \bibfield  {author} {\bibinfo {author} {\bibfnamefont {M.}~\bibnamefont {Shorbagy}}\ and\ \bibinfo {author} {\bibfnamefont {H.}~\bibnamefont {Taha}},\ }\bibfield  {title} {\enquote {\bibinfo {title} {Magnus force estimation using gauss’s principle of least constraint},}\ }\href@noop {} {\bibfield  {journal} {\bibinfo  {journal} {AIAA Journal}\ }\textbf {\bibinfo {volume} {62}},\ \bibinfo {pages} {1962--1969} (\bibinfo {year} {2024})}\BibitemShut {NoStop}%
\bibitem [{\citenamefont {Moin}\ and\ \citenamefont {Kim}(1980)}]{moin1980numerical}%
  \BibitemOpen
  \bibfield  {author} {\bibinfo {author} {\bibfnamefont {P.}~\bibnamefont {Moin}}\ and\ \bibinfo {author} {\bibfnamefont {J.}~\bibnamefont {Kim}},\ }\bibfield  {title} {\enquote {\bibinfo {title} {On the numerical solution of time-dependent viscous incompressible fluid flows involving solid boundaries},}\ }\href@noop {} {\bibfield  {journal} {\bibinfo  {journal} {Journal of computational physics}\ }\textbf {\bibinfo {volume} {35}},\ \bibinfo {pages} {381--392} (\bibinfo {year} {1980})}\BibitemShut {NoStop}%
\bibitem [{\citenamefont {Gresho}\ and\ \citenamefont {Sani}(1987)}]{gresho1987pressure}%
  \BibitemOpen
  \bibfield  {author} {\bibinfo {author} {\bibfnamefont {P.~M.}\ \bibnamefont {Gresho}}\ and\ \bibinfo {author} {\bibfnamefont {R.~L.}\ \bibnamefont {Sani}},\ }\bibfield  {title} {\enquote {\bibinfo {title} {On pressure boundary conditions for the incompressible navier-stokes equations},}\ }\href@noop {} {\bibfield  {journal} {\bibinfo  {journal} {International Journal for Numerical Methods in Fluids}\ }\textbf {\bibinfo {volume} {7}},\ \bibinfo {pages} {1111--1145} (\bibinfo {year} {1987})}\BibitemShut {NoStop}%
\bibitem [{\citenamefont {Chorin}, \citenamefont {Marsden},\ and\ \citenamefont {Marsden}(1990)}]{chorin1990mathematical}%
  \BibitemOpen
  \bibfield  {author} {\bibinfo {author} {\bibfnamefont {A.~J.}\ \bibnamefont {Chorin}}, \bibinfo {author} {\bibfnamefont {J.~E.}\ \bibnamefont {Marsden}}, \ and\ \bibinfo {author} {\bibfnamefont {J.~E.}\ \bibnamefont {Marsden}},\ }\href@noop {} {\emph {\bibinfo {title} {A mathematical introduction to fluid mechanics}}},\ Vol.~\bibinfo {volume} {3}\ (\bibinfo  {publisher} {Springer},\ \bibinfo {year} {1990})\BibitemShut {NoStop}%
\bibitem [{\citenamefont {Kambe}(2009)}]{kambe2009geometrical}%
  \BibitemOpen
  \bibfield  {author} {\bibinfo {author} {\bibfnamefont {T.~J.}\ \bibnamefont {Kambe}},\ }\href@noop {} {\emph {\bibinfo {title} {Geometrical Theory Of Dynamical Systems And Fluid Flows (Revised Edition)}}},\ Vol.~\bibinfo {volume} {23}\ (\bibinfo  {publisher} {World Scientific},\ \bibinfo {year} {2009})\BibitemShut {NoStop}%
\bibitem [{\citenamefont {Arnold}(1966)}]{arnold1966geometrie}%
  \BibitemOpen
  \bibfield  {author} {\bibinfo {author} {\bibfnamefont {V.}~\bibnamefont {Arnold}},\ }\bibfield  {title} {\enquote {\bibinfo {title} {Sur la g{\'e}om{\'e}trie diff{\'e}rentielle des groupes de lie de dimension infinie et ses applications {\`a} l'hydrodynamique des fluides parfaits},}\ }in\ \href@noop {} {\emph {\bibinfo {booktitle} {Annales de l'institut Fourier}}},\ Vol.~\bibinfo {volume} {16}\ (\bibinfo {year} {1966})\ pp.\ \bibinfo {pages} {319--361}\BibitemShut {NoStop}%
\bibitem [{\citenamefont {Krizhevsky}, \citenamefont {Sutskever},\ and\ \citenamefont {Hinton}(2012)}]{krizhevsky2012imagenet}%
  \BibitemOpen
  \bibfield  {author} {\bibinfo {author} {\bibfnamefont {A.}~\bibnamefont {Krizhevsky}}, \bibinfo {author} {\bibfnamefont {I.}~\bibnamefont {Sutskever}}, \ and\ \bibinfo {author} {\bibfnamefont {G.~E.}\ \bibnamefont {Hinton}},\ }\bibfield  {title} {\enquote {\bibinfo {title} {Imagenet classification with deep convolutional neural networks},}\ }\href@noop {} {\bibfield  {journal} {\bibinfo  {journal} {Advances in neural information processing systems}\ }\textbf {\bibinfo {volume} {25}},\ \bibinfo {pages} {1097--1105} (\bibinfo {year} {2012})}\BibitemShut {NoStop}%
\bibitem [{\citenamefont {He}\ \emph {et~al.}(2016)\citenamefont {He}, \citenamefont {Zhang}, \citenamefont {Ren},\ and\ \citenamefont {Sun}}]{he2016deep}%
  \BibitemOpen
  \bibfield  {author} {\bibinfo {author} {\bibfnamefont {K.}~\bibnamefont {He}}, \bibinfo {author} {\bibfnamefont {X.}~\bibnamefont {Zhang}}, \bibinfo {author} {\bibfnamefont {S.}~\bibnamefont {Ren}}, \ and\ \bibinfo {author} {\bibfnamefont {J.}~\bibnamefont {Sun}},\ }\bibfield  {title} {\enquote {\bibinfo {title} {Deep residual learning for image recognition},}\ }\href@noop {} {\bibfield  {journal} {\bibinfo  {journal} {Proceedings of the IEEE conference on computer vision and pattern recognition}\ ,\ \bibinfo {pages} {770--778}} (\bibinfo {year} {2016})}\BibitemShut {NoStop}%
\bibitem [{\citenamefont {Redmon}\ \emph {et~al.}(2016)\citenamefont {Redmon}, \citenamefont {Divvala}, \citenamefont {Girshick},\ and\ \citenamefont {Farhadi}}]{redmon2016you}%
  \BibitemOpen
  \bibfield  {author} {\bibinfo {author} {\bibfnamefont {J.}~\bibnamefont {Redmon}}, \bibinfo {author} {\bibfnamefont {S.}~\bibnamefont {Divvala}}, \bibinfo {author} {\bibfnamefont {R.}~\bibnamefont {Girshick}}, \ and\ \bibinfo {author} {\bibfnamefont {A.}~\bibnamefont {Farhadi}},\ }\bibfield  {title} {\enquote {\bibinfo {title} {You only look once: Unified, real-time object detection},}\ }\href@noop {} {\bibfield  {journal} {\bibinfo  {journal} {Proceedings of the IEEE conference on computer vision and pattern recognition}\ ,\ \bibinfo {pages} {779--788}} (\bibinfo {year} {2016})}\BibitemShut {NoStop}%
\bibitem [{\citenamefont {Covington}, \citenamefont {Adams},\ and\ \citenamefont {Sargin}(2016)}]{covington2016deep}%
  \BibitemOpen
  \bibfield  {author} {\bibinfo {author} {\bibfnamefont {P.}~\bibnamefont {Covington}}, \bibinfo {author} {\bibfnamefont {J.}~\bibnamefont {Adams}}, \ and\ \bibinfo {author} {\bibfnamefont {E.}~\bibnamefont {Sargin}},\ }\bibfield  {title} {\enquote {\bibinfo {title} {Deep neural networks for youtube recommendations},}\ }\href@noop {} {\bibfield  {journal} {\bibinfo  {journal} {Proceedings of the 10th ACM Conference on Recommender Systems}\ ,\ \bibinfo {pages} {191--198}} (\bibinfo {year} {2016})}\BibitemShut {NoStop}%
\bibitem [{\citenamefont {He}\ \emph {et~al.}(2017)\citenamefont {He}, \citenamefont {Liao}, \citenamefont {Zhang}, \citenamefont {Nie}, \citenamefont {Hu},\ and\ \citenamefont {Chua}}]{he2017neural}%
  \BibitemOpen
  \bibfield  {author} {\bibinfo {author} {\bibfnamefont {X.}~\bibnamefont {He}}, \bibinfo {author} {\bibfnamefont {L.}~\bibnamefont {Liao}}, \bibinfo {author} {\bibfnamefont {H.}~\bibnamefont {Zhang}}, \bibinfo {author} {\bibfnamefont {L.}~\bibnamefont {Nie}}, \bibinfo {author} {\bibfnamefont {X.}~\bibnamefont {Hu}}, \ and\ \bibinfo {author} {\bibfnamefont {T.-S.}\ \bibnamefont {Chua}},\ }\bibfield  {title} {\enquote {\bibinfo {title} {Neural collaborative filtering},}\ }\href@noop {} {\bibfield  {journal} {\bibinfo  {journal} {Proceedings of the 26th International Conference on World Wide Web}\ ,\ \bibinfo {pages} {173--182}} (\bibinfo {year} {2017})}\BibitemShut {NoStop}%
\bibitem [{\citenamefont {Esteva}\ \emph {et~al.}(2017)\citenamefont {Esteva}, \citenamefont {Kuprel}, \citenamefont {Novoa}, \citenamefont {Ko}, \citenamefont {Swetter}, \citenamefont {Blau},\ and\ \citenamefont {Thrun}}]{esteva2017dermatologist}%
  \BibitemOpen
  \bibfield  {author} {\bibinfo {author} {\bibfnamefont {A.}~\bibnamefont {Esteva}}, \bibinfo {author} {\bibfnamefont {B.}~\bibnamefont {Kuprel}}, \bibinfo {author} {\bibfnamefont {R.~A.}\ \bibnamefont {Novoa}}, \bibinfo {author} {\bibfnamefont {J.}~\bibnamefont {Ko}}, \bibinfo {author} {\bibfnamefont {S.~M.}\ \bibnamefont {Swetter}}, \bibinfo {author} {\bibfnamefont {H.~M.}\ \bibnamefont {Blau}}, \ and\ \bibinfo {author} {\bibfnamefont {S.}~\bibnamefont {Thrun}},\ }\bibfield  {title} {\enquote {\bibinfo {title} {Dermatologist-level classification of skin cancer with deep neural networks},}\ }\href@noop {} {\bibfield  {journal} {\bibinfo  {journal} {Nature}\ }\textbf {\bibinfo {volume} {542}},\ \bibinfo {pages} {115--118} (\bibinfo {year} {2017})}\BibitemShut {NoStop}%
\bibitem [{\citenamefont {Shen}\ \emph {et~al.}(2017)\citenamefont {Shen}, \citenamefont {Lin}, \citenamefont {Qing},\ and\ \citenamefont {Okamoto}}]{shen2017deep}%
  \BibitemOpen
  \bibfield  {author} {\bibinfo {author} {\bibfnamefont {Z.}~\bibnamefont {Shen}}, \bibinfo {author} {\bibfnamefont {Z.}~\bibnamefont {Lin}}, \bibinfo {author} {\bibfnamefont {C.}~\bibnamefont {Qing}}, \ and\ \bibinfo {author} {\bibfnamefont {E.}~\bibnamefont {Okamoto}},\ }\bibfield  {title} {\enquote {\bibinfo {title} {Deep residual learning for image steganalysis},}\ }\href@noop {} {\bibfield  {journal} {\bibinfo  {journal} {IEEE Transactions on Information Forensics and Security}\ }\textbf {\bibinfo {volume} {12}},\ \bibinfo {pages} {2545--2557} (\bibinfo {year} {2017})}\BibitemShut {NoStop}%
\bibitem [{\citenamefont {Tsantekidis}\ \emph {et~al.}(2017)\citenamefont {Tsantekidis}, \citenamefont {Passalis}, \citenamefont {Tefas}, \citenamefont {Kanniainen}, \citenamefont {Gabbouj},\ and\ \citenamefont {Iosifidis}}]{tsantekidis2017forecasting}%
  \BibitemOpen
  \bibfield  {author} {\bibinfo {author} {\bibfnamefont {A.}~\bibnamefont {Tsantekidis}}, \bibinfo {author} {\bibfnamefont {N.}~\bibnamefont {Passalis}}, \bibinfo {author} {\bibfnamefont {A.}~\bibnamefont {Tefas}}, \bibinfo {author} {\bibfnamefont {J.}~\bibnamefont {Kanniainen}}, \bibinfo {author} {\bibfnamefont {M.}~\bibnamefont {Gabbouj}}, \ and\ \bibinfo {author} {\bibfnamefont {A.}~\bibnamefont {Iosifidis}},\ }\bibfield  {title} {\enquote {\bibinfo {title} {Forecasting stock prices from the limit order book using convolutional neural networks},}\ }\href@noop {} {\bibfield  {journal} {\bibinfo  {journal} {IEEE Transactions on Neural Networks and Learning Systems}\ }\textbf {\bibinfo {volume} {29}},\ \bibinfo {pages} {5440--5449} (\bibinfo {year} {2017})}\BibitemShut {NoStop}%
\bibitem [{\citenamefont {Ding}\ \emph {et~al.}(2016)\citenamefont {Ding}, \citenamefont {Zhang}, \citenamefont {Liu}, \citenamefont {Duan},\ and\ \citenamefont {Ma}}]{ding2015deep}%
  \BibitemOpen
  \bibfield  {author} {\bibinfo {author} {\bibfnamefont {X.}~\bibnamefont {Ding}}, \bibinfo {author} {\bibfnamefont {Y.}~\bibnamefont {Zhang}}, \bibinfo {author} {\bibfnamefont {T.}~\bibnamefont {Liu}}, \bibinfo {author} {\bibfnamefont {J.}~\bibnamefont {Duan}}, \ and\ \bibinfo {author} {\bibfnamefont {Z.}~\bibnamefont {Ma}},\ }\bibfield  {title} {\enquote {\bibinfo {title} {Deep learning for event-driven stock prediction},}\ }\href@noop {} {\bibfield  {journal} {\bibinfo  {journal} {Neural Networks}\ }\textbf {\bibinfo {volume} {78}},\ \bibinfo {pages} {1--9} (\bibinfo {year} {2016})}\BibitemShut {NoStop}%
\bibitem [{\citenamefont {Shi}\ \emph {et~al.}(2015)\citenamefont {Shi}, \citenamefont {Gao}, \citenamefont {Lausen}, \citenamefont {Wang}, \citenamefont {Yeung}, \citenamefont {Wong},\ and\ \citenamefont {Woo}}]{shi2015convolutional}%
  \BibitemOpen
  \bibfield  {author} {\bibinfo {author} {\bibfnamefont {X.}~\bibnamefont {Shi}}, \bibinfo {author} {\bibfnamefont {Z.}~\bibnamefont {Gao}}, \bibinfo {author} {\bibfnamefont {L.}~\bibnamefont {Lausen}}, \bibinfo {author} {\bibfnamefont {H.}~\bibnamefont {Wang}}, \bibinfo {author} {\bibfnamefont {D.-Y.}\ \bibnamefont {Yeung}}, \bibinfo {author} {\bibfnamefont {W.-K.}\ \bibnamefont {Wong}}, \ and\ \bibinfo {author} {\bibfnamefont {W.-c.}\ \bibnamefont {Woo}},\ }\bibfield  {title} {\enquote {\bibinfo {title} {Convolutional lstm network: A machine learning approach for precipitation nowcasting},}\ }\href@noop {} {\bibfield  {journal} {\bibinfo  {journal} {Advances in neural information processing systems}\ ,\ \bibinfo {pages} {802--810}} (\bibinfo {year} {2015})}\BibitemShut {NoStop}%
\bibitem [{\citenamefont {Lipton}\ \emph {et~al.}(2015)\citenamefont {Lipton}, \citenamefont {Kale}, \citenamefont {Elkan},\ and\ \citenamefont {Wetzel}}]{lipton2015learning}%
  \BibitemOpen
  \bibfield  {author} {\bibinfo {author} {\bibfnamefont {Z.~C.}\ \bibnamefont {Lipton}}, \bibinfo {author} {\bibfnamefont {D.}~\bibnamefont {Kale}}, \bibinfo {author} {\bibfnamefont {C.}~\bibnamefont {Elkan}}, \ and\ \bibinfo {author} {\bibfnamefont {R.}~\bibnamefont {Wetzel}},\ }\bibfield  {title} {\enquote {\bibinfo {title} {Learning to diagnose with lstm recurrent neural networks},}\ }\href@noop {} {\bibfield  {journal} {\bibinfo  {journal} {arXiv preprint arXiv:1511.03677}\ } (\bibinfo {year} {2015})}\BibitemShut {NoStop}%
\bibitem [{\citenamefont {Silver}\ \emph {et~al.}(2016)\citenamefont {Silver}, \citenamefont {Huang}, \citenamefont {Maddison}, \citenamefont {Guez}, \citenamefont {Sifre}, \citenamefont {van~den Driessche}, \citenamefont {Schrittwieser}, \citenamefont {Antonoglou}, \citenamefont {Panneershelvam}, \citenamefont {Lanctot} \emph {et~al.}}]{silver2016mastering}%
  \BibitemOpen
  \bibfield  {author} {\bibinfo {author} {\bibfnamefont {D.}~\bibnamefont {Silver}}, \bibinfo {author} {\bibfnamefont {A.}~\bibnamefont {Huang}}, \bibinfo {author} {\bibfnamefont {C.~J.}\ \bibnamefont {Maddison}}, \bibinfo {author} {\bibfnamefont {A.}~\bibnamefont {Guez}}, \bibinfo {author} {\bibfnamefont {L.}~\bibnamefont {Sifre}}, \bibinfo {author} {\bibfnamefont {G.}~\bibnamefont {van~den Driessche}}, \bibinfo {author} {\bibfnamefont {J.}~\bibnamefont {Schrittwieser}}, \bibinfo {author} {\bibfnamefont {I.}~\bibnamefont {Antonoglou}}, \bibinfo {author} {\bibfnamefont {V.}~\bibnamefont {Panneershelvam}}, \bibinfo {author} {\bibfnamefont {M.}~\bibnamefont {Lanctot}},  \emph {et~al.},\ }\bibfield  {title} {\enquote {\bibinfo {title} {Mastering the game of go with deep neural networks and tree search},}\ }\href@noop {} {\bibfield  {journal} {\bibinfo  {journal} {nature}\ }\textbf {\bibinfo {volume} {529}},\ \bibinfo {pages} {484--489} (\bibinfo {year} {2016})}\BibitemShut {NoStop}%
\bibitem [{\citenamefont {Lanctot}\ \emph {et~al.}(2017)\citenamefont {Lanctot}, \citenamefont {Zambaldi}, \citenamefont {Gruslys}, \citenamefont {Lazaridou}, \citenamefont {Perolat}, \citenamefont {Silver},\ and\ \citenamefont {Graepel}}]{lanctot2017unified}%
  \BibitemOpen
  \bibfield  {author} {\bibinfo {author} {\bibfnamefont {M.}~\bibnamefont {Lanctot}}, \bibinfo {author} {\bibfnamefont {V.}~\bibnamefont {Zambaldi}}, \bibinfo {author} {\bibfnamefont {A.}~\bibnamefont {Gruslys}}, \bibinfo {author} {\bibfnamefont {A.}~\bibnamefont {Lazaridou}}, \bibinfo {author} {\bibfnamefont {J.}~\bibnamefont {Perolat}}, \bibinfo {author} {\bibfnamefont {D.}~\bibnamefont {Silver}}, \ and\ \bibinfo {author} {\bibfnamefont {T.}~\bibnamefont {Graepel}},\ }\bibfield  {title} {\enquote {\bibinfo {title} {A unified game-theoretic approach to multiagent reinforcement learning},}\ }\href@noop {} {\bibfield  {journal} {\bibinfo  {journal} {Advances in neural information processing systems}\ ,\ \bibinfo {pages} {4190--4203}} (\bibinfo {year} {2017})}\BibitemShut {NoStop}%
\bibitem [{\citenamefont {Mnih}\ \emph {et~al.}(2015)\citenamefont {Mnih}, \citenamefont {Kavukcuoglu}, \citenamefont {Silver}, \citenamefont {Rusu}, \citenamefont {Veness}, \citenamefont {Bellemare}, \citenamefont {Graves}, \citenamefont {Riedmiller}, \citenamefont {Fidjeland}, \citenamefont {Ostrovski} \emph {et~al.}}]{mnih2015human}%
  \BibitemOpen
  \bibfield  {author} {\bibinfo {author} {\bibfnamefont {V.}~\bibnamefont {Mnih}}, \bibinfo {author} {\bibfnamefont {K.}~\bibnamefont {Kavukcuoglu}}, \bibinfo {author} {\bibfnamefont {D.}~\bibnamefont {Silver}}, \bibinfo {author} {\bibfnamefont {A.~A.}\ \bibnamefont {Rusu}}, \bibinfo {author} {\bibfnamefont {J.}~\bibnamefont {Veness}}, \bibinfo {author} {\bibfnamefont {M.~G.}\ \bibnamefont {Bellemare}}, \bibinfo {author} {\bibfnamefont {A.}~\bibnamefont {Graves}}, \bibinfo {author} {\bibfnamefont {M.}~\bibnamefont {Riedmiller}}, \bibinfo {author} {\bibfnamefont {A.~K.}\ \bibnamefont {Fidjeland}}, \bibinfo {author} {\bibfnamefont {G.}~\bibnamefont {Ostrovski}},  \emph {et~al.},\ }\bibfield  {title} {\enquote {\bibinfo {title} {Human-level control through deep reinforcement learning},}\ }\href@noop {} {\bibfield  {journal} {\bibinfo  {journal} {Nature}\ }\textbf {\bibinfo {volume} {518}},\ \bibinfo {pages} {529--533} (\bibinfo {year} {2015})}\BibitemShut {NoStop}%
\bibitem [{\citenamefont {Hassabis}\ \emph {et~al.}(2017)\citenamefont {Hassabis}, \citenamefont {Kumaran}, \citenamefont {Summerfield},\ and\ \citenamefont {Botvinick}}]{hassabis2017neuroscience}%
  \BibitemOpen
  \bibfield  {author} {\bibinfo {author} {\bibfnamefont {D.}~\bibnamefont {Hassabis}}, \bibinfo {author} {\bibfnamefont {D.}~\bibnamefont {Kumaran}}, \bibinfo {author} {\bibfnamefont {C.}~\bibnamefont {Summerfield}}, \ and\ \bibinfo {author} {\bibfnamefont {M.}~\bibnamefont {Botvinick}},\ }\bibfield  {title} {\enquote {\bibinfo {title} {Neuroscience-inspired artificial intelligence},}\ }\href@noop {} {\bibfield  {journal} {\bibinfo  {journal} {Neuron}\ }\textbf {\bibinfo {volume} {95}},\ \bibinfo {pages} {245--258} (\bibinfo {year} {2017})}\BibitemShut {NoStop}%
\bibitem [{\citenamefont {Angermueller}\ \emph {et~al.}(2016)\citenamefont {Angermueller}, \citenamefont {P{\"a}rnamaa}, \citenamefont {Parts},\ and\ \citenamefont {Stegle}}]{angermueller2016deep}%
  \BibitemOpen
  \bibfield  {author} {\bibinfo {author} {\bibfnamefont {C.}~\bibnamefont {Angermueller}}, \bibinfo {author} {\bibfnamefont {T.}~\bibnamefont {P{\"a}rnamaa}}, \bibinfo {author} {\bibfnamefont {L.}~\bibnamefont {Parts}}, \ and\ \bibinfo {author} {\bibfnamefont {O.}~\bibnamefont {Stegle}},\ }\bibfield  {title} {\enquote {\bibinfo {title} {Deep learning for computational biology},}\ }\href@noop {} {\bibfield  {journal} {\bibinfo  {journal} {Molecular Systems Biology}\ }\textbf {\bibinfo {volume} {12}},\ \bibinfo {pages} {878} (\bibinfo {year} {2016})}\BibitemShut {NoStop}%
\bibitem [{\citenamefont {Zhou}\ \emph {et~al.}(2015)\citenamefont {Zhou}, \citenamefont {Yu}, \citenamefont {Du}, \citenamefont {Yao}, \citenamefont {Li}, \citenamefont {Huang},\ and\ \citenamefont {Wu}}]{zhou2015predicting}%
  \BibitemOpen
  \bibfield  {author} {\bibinfo {author} {\bibfnamefont {J.}~\bibnamefont {Zhou}}, \bibinfo {author} {\bibfnamefont {W.}~\bibnamefont {Yu}}, \bibinfo {author} {\bibfnamefont {Y.}~\bibnamefont {Du}}, \bibinfo {author} {\bibfnamefont {X.}~\bibnamefont {Yao}}, \bibinfo {author} {\bibfnamefont {L.}~\bibnamefont {Li}}, \bibinfo {author} {\bibfnamefont {H.}~\bibnamefont {Huang}}, \ and\ \bibinfo {author} {\bibfnamefont {B.}~\bibnamefont {Wu}},\ }\bibfield  {title} {\enquote {\bibinfo {title} {Predicting comprehensive cancer vulnerability from genomic and epigenomic features},}\ }\href@noop {} {\bibfield  {journal} {\bibinfo  {journal} {Bioinformatics}\ }\textbf {\bibinfo {volume} {31}},\ \bibinfo {pages} {i222--i230} (\bibinfo {year} {2015})}\BibitemShut {NoStop}%
\bibitem [{\citenamefont {Bojarski}\ \emph {et~al.}(2016)\citenamefont {Bojarski}, \citenamefont {Del~Testa}, \citenamefont {Dworakowski}, \citenamefont {Firner}, \citenamefont {Flepp}, \citenamefont {Goyal}, \citenamefont {Jackel}, \citenamefont {Monfort}, \citenamefont {Muller}, \citenamefont {Zhang} \emph {et~al.}}]{bojarski2016end}%
  \BibitemOpen
  \bibfield  {author} {\bibinfo {author} {\bibfnamefont {M.}~\bibnamefont {Bojarski}}, \bibinfo {author} {\bibfnamefont {D.}~\bibnamefont {Del~Testa}}, \bibinfo {author} {\bibfnamefont {D.}~\bibnamefont {Dworakowski}}, \bibinfo {author} {\bibfnamefont {B.}~\bibnamefont {Firner}}, \bibinfo {author} {\bibfnamefont {B.}~\bibnamefont {Flepp}}, \bibinfo {author} {\bibfnamefont {P.}~\bibnamefont {Goyal}}, \bibinfo {author} {\bibfnamefont {L.~D.}\ \bibnamefont {Jackel}}, \bibinfo {author} {\bibfnamefont {M.}~\bibnamefont {Monfort}}, \bibinfo {author} {\bibfnamefont {U.}~\bibnamefont {Muller}}, \bibinfo {author} {\bibfnamefont {J.}~\bibnamefont {Zhang}},  \emph {et~al.},\ }\bibfield  {title} {\enquote {\bibinfo {title} {End to end learning for self-driving cars},}\ }\href@noop {} {\bibfield  {journal} {\bibinfo  {journal} {arXiv preprint arXiv:1604.07316}\ } (\bibinfo {year} {2016})}\BibitemShut {NoStop}%
\bibitem [{\citenamefont {Chen}\ \emph {et~al.}()\citenamefont {Chen}, \citenamefont {Seff}, \citenamefont {Kornhauser},\ and\ \citenamefont {Xiao}}]{chen2015deepdriving}%
  \BibitemOpen
  \bibfield  {author} {\bibinfo {author} {\bibfnamefont {T.}~\bibnamefont {Chen}}, \bibinfo {author} {\bibfnamefont {A.}~\bibnamefont {Seff}}, \bibinfo {author} {\bibfnamefont {A.}~\bibnamefont {Kornhauser}}, \ and\ \bibinfo {author} {\bibfnamefont {J.}~\bibnamefont {Xiao}},\ }\bibfield  {title} {\enquote {\bibinfo {title} {Deepdriving: Learning affordance for direct perception in autonomous driving},}\ }\href@noop {} {\bibinfo  {journal} {arXiv preprint arXiv:1505.00256}\ }\BibitemShut {NoStop}%
\bibitem [{\citenamefont {Silver}\ \emph {et~al.}(2017)\citenamefont {Silver}, \citenamefont {Schrittwieser}, \citenamefont {Simonyan}, \citenamefont {Antonoglou}, \citenamefont {Huang}, \citenamefont {Guez}, \citenamefont {Hubert}, \citenamefont {Baker}, \citenamefont {Lai}, \citenamefont {Bolton} \emph {et~al.}}]{silver2017mastering}%
  \BibitemOpen
\bibfield  {journal} {  }\bibfield  {author} {\bibinfo {author} {\bibfnamefont {D.}~\bibnamefont {Silver}}, \bibinfo {author} {\bibfnamefont {J.}~\bibnamefont {Schrittwieser}}, \bibinfo {author} {\bibfnamefont {K.}~\bibnamefont {Simonyan}}, \bibinfo {author} {\bibfnamefont {I.}~\bibnamefont {Antonoglou}}, \bibinfo {author} {\bibfnamefont {A.}~\bibnamefont {Huang}}, \bibinfo {author} {\bibfnamefont {A.}~\bibnamefont {Guez}}, \bibinfo {author} {\bibfnamefont {T.}~\bibnamefont {Hubert}}, \bibinfo {author} {\bibfnamefont {L.}~\bibnamefont {Baker}}, \bibinfo {author} {\bibfnamefont {M.}~\bibnamefont {Lai}}, \bibinfo {author} {\bibfnamefont {A.}~\bibnamefont {Bolton}},  \emph {et~al.},\ }\bibfield  {title} {\enquote {\bibinfo {title} {Mastering the game of go without human knowledge},}\ }\href@noop {} {\bibfield  {journal} {\bibinfo  {journal} {nature}\ }\textbf {\bibinfo {volume} {550}},\ \bibinfo {pages} {354--359} (\bibinfo {year} {2017})}\BibitemShut {NoStop}%
\bibitem [{\citenamefont {Hinton}\ \emph {et~al.}(2012)\citenamefont {Hinton}, \citenamefont {Deng}, \citenamefont {Yu}, \citenamefont {Dahl}, \citenamefont {Mohamed}, \citenamefont {Jaitly}, \citenamefont {Senior}, \citenamefont {Vanhoucke}, \citenamefont {Nguyen}, \citenamefont {Sainath} \emph {et~al.}}]{hinton2012deep}%
  \BibitemOpen
  \bibfield  {author} {\bibinfo {author} {\bibfnamefont {G.}~\bibnamefont {Hinton}}, \bibinfo {author} {\bibfnamefont {L.}~\bibnamefont {Deng}}, \bibinfo {author} {\bibfnamefont {D.}~\bibnamefont {Yu}}, \bibinfo {author} {\bibfnamefont {G.~E.}\ \bibnamefont {Dahl}}, \bibinfo {author} {\bibfnamefont {A.-r.}\ \bibnamefont {Mohamed}}, \bibinfo {author} {\bibfnamefont {N.}~\bibnamefont {Jaitly}}, \bibinfo {author} {\bibfnamefont {A.}~\bibnamefont {Senior}}, \bibinfo {author} {\bibfnamefont {V.}~\bibnamefont {Vanhoucke}}, \bibinfo {author} {\bibfnamefont {P.}~\bibnamefont {Nguyen}}, \bibinfo {author} {\bibfnamefont {T.~N.}\ \bibnamefont {Sainath}},  \emph {et~al.},\ }\bibfield  {title} {\enquote {\bibinfo {title} {Deep neural networks for acoustic modeling in speech recognition: The shared views of four research groups},}\ }\href@noop {} {\bibfield  {journal} {\bibinfo  {journal} {IEEE Signal Processing Magazine}\ }\textbf {\bibinfo {volume} {29}},\ \bibinfo {pages} {82--97} (\bibinfo {year} {2012})}\BibitemShut
  {NoStop}%
\bibitem [{\citenamefont {Bahdanau}, \citenamefont {Cho},\ and\ \citenamefont {Bengio}(2014)}]{bahdanau2014neural}%
  \BibitemOpen
  \bibfield  {author} {\bibinfo {author} {\bibfnamefont {D.}~\bibnamefont {Bahdanau}}, \bibinfo {author} {\bibfnamefont {K.}~\bibnamefont {Cho}}, \ and\ \bibinfo {author} {\bibfnamefont {Y.}~\bibnamefont {Bengio}},\ }\bibfield  {title} {\enquote {\bibinfo {title} {Neural machine translation by jointly learning to align and translate},}\ }\href@noop {} {\bibfield  {journal} {\bibinfo  {journal} {arXiv preprint arXiv:1409.0473}\ } (\bibinfo {year} {2014})}\BibitemShut {NoStop}%
\bibitem [{\citenamefont {Vaswani}\ \emph {et~al.}(2017)\citenamefont {Vaswani}, \citenamefont {Shazeer}, \citenamefont {Parmar}, \citenamefont {Uszkoreit}, \citenamefont {Jones}, \citenamefont {Gomez}, \citenamefont {Kaiser},\ and\ \citenamefont {Polosukhin}}]{vaswani2017attention}%
  \BibitemOpen
  \bibfield  {author} {\bibinfo {author} {\bibfnamefont {A.}~\bibnamefont {Vaswani}}, \bibinfo {author} {\bibfnamefont {N.}~\bibnamefont {Shazeer}}, \bibinfo {author} {\bibfnamefont {N.}~\bibnamefont {Parmar}}, \bibinfo {author} {\bibfnamefont {J.}~\bibnamefont {Uszkoreit}}, \bibinfo {author} {\bibfnamefont {L.}~\bibnamefont {Jones}}, \bibinfo {author} {\bibfnamefont {A.~N.}\ \bibnamefont {Gomez}}, \bibinfo {author} {\bibfnamefont {{\L}.}~\bibnamefont {Kaiser}}, \ and\ \bibinfo {author} {\bibfnamefont {I.}~\bibnamefont {Polosukhin}},\ }\bibfield  {title} {\enquote {\bibinfo {title} {Attention is all you need},}\ }\href@noop {} {\bibfield  {journal} {\bibinfo  {journal} {Advances in neural information processing systems}\ ,\ \bibinfo {pages} {5998--6008}} (\bibinfo {year} {2017})}\BibitemShut {NoStop}%
\bibitem [{\citenamefont {Wu}\ \emph {et~al.}(2023)\citenamefont {Wu}, \citenamefont {He}, \citenamefont {Liu}, \citenamefont {Sun}, \citenamefont {Liu}, \citenamefont {Han},\ and\ \citenamefont {Tang}}]{wu2023brief}%
  \BibitemOpen
  \bibfield  {author} {\bibinfo {author} {\bibfnamefont {T.}~\bibnamefont {Wu}}, \bibinfo {author} {\bibfnamefont {S.}~\bibnamefont {He}}, \bibinfo {author} {\bibfnamefont {J.}~\bibnamefont {Liu}}, \bibinfo {author} {\bibfnamefont {S.}~\bibnamefont {Sun}}, \bibinfo {author} {\bibfnamefont {K.}~\bibnamefont {Liu}}, \bibinfo {author} {\bibfnamefont {Q.-L.}\ \bibnamefont {Han}}, \ and\ \bibinfo {author} {\bibfnamefont {Y.}~\bibnamefont {Tang}},\ }\bibfield  {title} {\enquote {\bibinfo {title} {A brief overview of chatgpt: The history, status quo and potential future development},}\ }\href@noop {} {\bibfield  {journal} {\bibinfo  {journal} {IEEE/CAA Journal of Automatica Sinica}\ }\textbf {\bibinfo {volume} {10}},\ \bibinfo {pages} {1122--1136} (\bibinfo {year} {2023})}\BibitemShut {NoStop}%
\bibitem [{\citenamefont {Roumeliotis}\ and\ \citenamefont {Tselikas}(2023)}]{roumeliotis2023chatgpt}%
  \BibitemOpen
  \bibfield  {author} {\bibinfo {author} {\bibfnamefont {K.~I.}\ \bibnamefont {Roumeliotis}}\ and\ \bibinfo {author} {\bibfnamefont {N.~D.}\ \bibnamefont {Tselikas}},\ }\bibfield  {title} {\enquote {\bibinfo {title} {Chatgpt and open-ai models: A preliminary review},}\ }\href@noop {} {\bibfield  {journal} {\bibinfo  {journal} {Future Internet}\ }\textbf {\bibinfo {volume} {15}},\ \bibinfo {pages} {192} (\bibinfo {year} {2023})}\BibitemShut {NoStop}%
\bibitem [{\citenamefont {Lagaris}, \citenamefont {Likas},\ and\ \citenamefont {Fotiadis}(1998)}]{lagaris1998artificial}%
  \BibitemOpen
  \bibfield  {author} {\bibinfo {author} {\bibfnamefont {I.~E.}\ \bibnamefont {Lagaris}}, \bibinfo {author} {\bibfnamefont {A.}~\bibnamefont {Likas}}, \ and\ \bibinfo {author} {\bibfnamefont {D.~I.}\ \bibnamefont {Fotiadis}},\ }\bibfield  {title} {\enquote {\bibinfo {title} {Artificial neural networks for solving ordinary and partial differential equations},}\ }\href@noop {} {\bibfield  {journal} {\bibinfo  {journal} {IEEE transactions on neural networks}\ }\textbf {\bibinfo {volume} {9}},\ \bibinfo {pages} {987--1000} (\bibinfo {year} {1998})}\BibitemShut {NoStop}%
\bibitem [{\citenamefont {Lagaris}, \citenamefont {Likas},\ and\ \citenamefont {Papageorgiou}(2000)}]{lagaris2000neural}%
  \BibitemOpen
  \bibfield  {author} {\bibinfo {author} {\bibfnamefont {I.~E.}\ \bibnamefont {Lagaris}}, \bibinfo {author} {\bibfnamefont {A.~C.}\ \bibnamefont {Likas}}, \ and\ \bibinfo {author} {\bibfnamefont {D.~G.}\ \bibnamefont {Papageorgiou}},\ }\bibfield  {title} {\enquote {\bibinfo {title} {Neural-network methods for boundary value problems with irregular boundaries},}\ }\href@noop {} {\bibfield  {journal} {\bibinfo  {journal} {IEEE Transactions on Neural Networks}\ }\textbf {\bibinfo {volume} {11}},\ \bibinfo {pages} {1041--1049} (\bibinfo {year} {2000})}\BibitemShut {NoStop}%
\bibitem [{\citenamefont {Owhadi}(2015)}]{owhadi2015bayesian}%
  \BibitemOpen
  \bibfield  {author} {\bibinfo {author} {\bibfnamefont {H.}~\bibnamefont {Owhadi}},\ }\bibfield  {title} {\enquote {\bibinfo {title} {Bayesian numerical homogenization},}\ }\href@noop {} {\bibfield  {journal} {\bibinfo  {journal} {Multiscale Modeling \& Simulation}\ }\textbf {\bibinfo {volume} {13}},\ \bibinfo {pages} {812--828} (\bibinfo {year} {2015})}\BibitemShut {NoStop}%
\bibitem [{\citenamefont {Raissi}, \citenamefont {Perdikaris},\ and\ \citenamefont {Karniadakis}(2017{\natexlab{a}})}]{raissi2017physics}%
  \BibitemOpen
  \bibfield  {author} {\bibinfo {author} {\bibfnamefont {M.}~\bibnamefont {Raissi}}, \bibinfo {author} {\bibfnamefont {P.}~\bibnamefont {Perdikaris}}, \ and\ \bibinfo {author} {\bibfnamefont {G.~E.}\ \bibnamefont {Karniadakis}},\ }\bibfield  {title} {\enquote {\bibinfo {title} {Physics informed deep learning (part i): Data-driven solutions of nonlinear partial differential equations},}\ }\href@noop {} {\bibfield  {journal} {\bibinfo  {journal} {arXiv preprint arXiv:1711.10561}\ } (\bibinfo {year} {2017}{\natexlab{a}})}\BibitemShut {NoStop}%
\bibitem [{\citenamefont {Raissi}, \citenamefont {Perdikaris},\ and\ \citenamefont {Karniadakis}(2017{\natexlab{b}})}]{raissi2017physics2}%
  \BibitemOpen
  \bibfield  {author} {\bibinfo {author} {\bibfnamefont {M.}~\bibnamefont {Raissi}}, \bibinfo {author} {\bibfnamefont {P.}~\bibnamefont {Perdikaris}}, \ and\ \bibinfo {author} {\bibfnamefont {G.~E.}\ \bibnamefont {Karniadakis}},\ }\bibfield  {title} {\enquote {\bibinfo {title} {Physics informed deep learning (part ii): Data-driven solutions of nonlinear partial differential equations},}\ }\href@noop {} {\bibfield  {journal} {\bibinfo  {journal} {arXiv preprint arXiv:1711.10566}\ } (\bibinfo {year} {2017}{\natexlab{b}})}\BibitemShut {NoStop}%
\bibitem [{\citenamefont {Cai}\ \emph {et~al.}(2021)\citenamefont {Cai}, \citenamefont {Mao}, \citenamefont {Wang}, \citenamefont {Yin},\ and\ \citenamefont {Karniadakis}}]{cai_physics-informed_2021}%
  \BibitemOpen
  \bibfield  {author} {\bibinfo {author} {\bibfnamefont {S.}~\bibnamefont {Cai}}, \bibinfo {author} {\bibfnamefont {Z.}~\bibnamefont {Mao}}, \bibinfo {author} {\bibfnamefont {Z.}~\bibnamefont {Wang}}, \bibinfo {author} {\bibfnamefont {M.}~\bibnamefont {Yin}}, \ and\ \bibinfo {author} {\bibfnamefont {G.~E.}\ \bibnamefont {Karniadakis}},\ }\bibfield  {title} {{\selectlanguage {en}\enquote {\bibinfo {title} {Physics-informed neural networks ({PINNs}) for fluid mechanics: a review},}\ }}\href {\doibase 10.1007/s10409-021-01148-1} {\bibfield  {journal} {\bibinfo  {journal} {Acta Mechanica Sinica}\ }\textbf {\bibinfo {volume} {37}},\ \bibinfo {pages} {1727--1738} (\bibinfo {year} {2021})}\BibitemShut {NoStop}%
\bibitem [{\citenamefont {Mao}, \citenamefont {Jagtap},\ and\ \citenamefont {Karniadakis}(2020)}]{mao_physics-informed_2020}%
  \BibitemOpen
  \bibfield  {author} {\bibinfo {author} {\bibfnamefont {Z.}~\bibnamefont {Mao}}, \bibinfo {author} {\bibfnamefont {A.~D.}\ \bibnamefont {Jagtap}}, \ and\ \bibinfo {author} {\bibfnamefont {G.~E.}\ \bibnamefont {Karniadakis}},\ }\bibfield  {title} {{\selectlanguage {en}\enquote {\bibinfo {title} {Physics-informed neural networks for high-speed flows},}\ }}\href {\doibase 10.1016/j.cma.2019.112789} {\bibfield  {journal} {\bibinfo  {journal} {Computer Methods in Applied Mechanics and Engineering}\ }\textbf {\bibinfo {volume} {360}},\ \bibinfo {pages} {112789} (\bibinfo {year} {2020})}\BibitemShut {NoStop}%
\bibitem [{\citenamefont {Blechschmidt}\ and\ \citenamefont {Ernst}(2021)}]{blechschmidt_three_2021}%
  \BibitemOpen
  \bibfield  {author} {\bibinfo {author} {\bibfnamefont {J.}~\bibnamefont {Blechschmidt}}\ and\ \bibinfo {author} {\bibfnamefont {O.~G.}\ \bibnamefont {Ernst}},\ }\bibfield  {title} {{\selectlanguage {en}\enquote {\bibinfo {title} {Three ways to solve partial differential equations with neural networks — {A} review},}\ }}\href {\doibase 10.1002/gamm.202100006} {\bibfield  {journal} {\bibinfo  {journal} {GAMM-Mitteilungen}\ }\textbf {\bibinfo {volume} {44}} (\bibinfo {year} {2021}),\ 10.1002/gamm.202100006}\BibitemShut {NoStop}%
\bibitem [{\citenamefont {Liu}\ and\ \citenamefont {Wang}(2019)}]{liu_multi-fidelity_2019}%
  \BibitemOpen
  \bibfield  {author} {\bibinfo {author} {\bibfnamefont {D.}~\bibnamefont {Liu}}\ and\ \bibinfo {author} {\bibfnamefont {Y.}~\bibnamefont {Wang}},\ }\bibfield  {title} {{\selectlanguage {en}\enquote {\bibinfo {title} {Multi-{Fidelity} {Physics}-{Constrained} {Neural} {Network} and {Its} {Application} in {Materials} {Modeling}},}\ }}\href {\doibase 10.1115/1.4044400} {\bibfield  {journal} {\bibinfo  {journal} {Journal of Mechanical Design}\ }\textbf {\bibinfo {volume} {141}},\ \bibinfo {pages} {121403} (\bibinfo {year} {2019})}\BibitemShut {NoStop}%
\bibitem [{\citenamefont {Nguyen-Thanh}, \citenamefont {Zhuang},\ and\ \citenamefont {Rabczuk}(2020)}]{nguyen-thanh_deep_2020}%
  \BibitemOpen
  \bibfield  {author} {\bibinfo {author} {\bibfnamefont {V.~M.}\ \bibnamefont {Nguyen-Thanh}}, \bibinfo {author} {\bibfnamefont {X.}~\bibnamefont {Zhuang}}, \ and\ \bibinfo {author} {\bibfnamefont {T.}~\bibnamefont {Rabczuk}},\ }\bibfield  {title} {{\selectlanguage {en}\enquote {\bibinfo {title} {A deep energy method for finite deformation hyperelasticity},}\ }}\href {\doibase 10.1016/j.euromechsol.2019.103874} {\bibfield  {journal} {\bibinfo  {journal} {European Journal of Mechanics - A/Solids}\ }\textbf {\bibinfo {volume} {80}},\ \bibinfo {pages} {103874} (\bibinfo {year} {2020})}\BibitemShut {NoStop}%
\bibitem [{\citenamefont {Goswami}\ \emph {et~al.}(2020)\citenamefont {Goswami}, \citenamefont {Anitescu}, \citenamefont {Chakraborty},\ and\ \citenamefont {Rabczuk}}]{goswami_transfer_2020}%
  \BibitemOpen
  \bibfield  {author} {\bibinfo {author} {\bibfnamefont {S.}~\bibnamefont {Goswami}}, \bibinfo {author} {\bibfnamefont {C.}~\bibnamefont {Anitescu}}, \bibinfo {author} {\bibfnamefont {S.}~\bibnamefont {Chakraborty}}, \ and\ \bibinfo {author} {\bibfnamefont {T.}~\bibnamefont {Rabczuk}},\ }\bibfield  {title} {{\selectlanguage {en}\enquote {\bibinfo {title} {Transfer learning enhanced physics informed neural network for phase-field modeling of fracture},}\ }}\href {\doibase 10.1016/j.tafmec.2019.102447} {\bibfield  {journal} {\bibinfo  {journal} {Theoretical and Applied Fracture Mechanics}\ }\textbf {\bibinfo {volume} {106}},\ \bibinfo {pages} {102447} (\bibinfo {year} {2020})}\BibitemShut {NoStop}%
\bibitem [{\citenamefont {Lu}\ \emph {et~al.}(2021)\citenamefont {Lu}, \citenamefont {Pestourie}, \citenamefont {Yao}, \citenamefont {Wang}, \citenamefont {Verdugo},\ and\ \citenamefont {Johnson}}]{lu2021physics}%
  \BibitemOpen
  \bibfield  {author} {\bibinfo {author} {\bibfnamefont {L.}~\bibnamefont {Lu}}, \bibinfo {author} {\bibfnamefont {R.}~\bibnamefont {Pestourie}}, \bibinfo {author} {\bibfnamefont {W.}~\bibnamefont {Yao}}, \bibinfo {author} {\bibfnamefont {Z.}~\bibnamefont {Wang}}, \bibinfo {author} {\bibfnamefont {F.}~\bibnamefont {Verdugo}}, \ and\ \bibinfo {author} {\bibfnamefont {S.~G.}\ \bibnamefont {Johnson}},\ }\bibfield  {title} {\enquote {\bibinfo {title} {Physics-informed neural networks with hard constraints for inverse design},}\ }\href@noop {} {\bibfield  {journal} {\bibinfo  {journal} {SIAM Journal on Scientific Computing}\ }\textbf {\bibinfo {volume} {43}},\ \bibinfo {pages} {B1105--B1132} (\bibinfo {year} {2021})}\BibitemShut {NoStop}%
\bibitem [{\citenamefont {Luenberger}, \citenamefont {Ye}\ \emph {et~al.}(1984)\citenamefont {Luenberger}, \citenamefont {Ye} \emph {et~al.}}]{luenberger1984linear}%
  \BibitemOpen
  \bibfield  {author} {\bibinfo {author} {\bibfnamefont {D.~G.}\ \bibnamefont {Luenberger}}, \bibinfo {author} {\bibfnamefont {Y.}~\bibnamefont {Ye}},  \emph {et~al.},\ }\href@noop {} {\emph {\bibinfo {title} {Linear and nonlinear programming}}},\ Vol.~\bibinfo {volume} {2}\ (\bibinfo  {publisher} {Springer},\ \bibinfo {year} {1984})\BibitemShut {NoStop}%
\bibitem [{\citenamefont {Long}\ \emph {et~al.}(2018)\citenamefont {Long}, \citenamefont {Lu}, \citenamefont {Ma},\ and\ \citenamefont {Dong}}]{long2018pde}%
  \BibitemOpen
  \bibfield  {author} {\bibinfo {author} {\bibfnamefont {Z.}~\bibnamefont {Long}}, \bibinfo {author} {\bibfnamefont {Y.}~\bibnamefont {Lu}}, \bibinfo {author} {\bibfnamefont {X.}~\bibnamefont {Ma}}, \ and\ \bibinfo {author} {\bibfnamefont {B.}~\bibnamefont {Dong}},\ }\bibfield  {title} {\enquote {\bibinfo {title} {Pde-net: Learning pdes from data},}\ }in\ \href@noop {} {\emph {\bibinfo {booktitle} {International conference on machine learning}}}\ (\bibinfo {organization} {PMLR},\ \bibinfo {year} {2018})\ pp.\ \bibinfo {pages} {3208--3216}\BibitemShut {NoStop}%
\bibitem [{\citenamefont {Hendriks}\ \emph {et~al.}(2020)\citenamefont {Hendriks}, \citenamefont {Jidling}, \citenamefont {Wills},\ and\ \citenamefont {Sch{\"o}n}}]{hendriks2020linearly}%
  \BibitemOpen
  \bibfield  {author} {\bibinfo {author} {\bibfnamefont {J.}~\bibnamefont {Hendriks}}, \bibinfo {author} {\bibfnamefont {C.}~\bibnamefont {Jidling}}, \bibinfo {author} {\bibfnamefont {A.}~\bibnamefont {Wills}}, \ and\ \bibinfo {author} {\bibfnamefont {T.}~\bibnamefont {Sch{\"o}n}},\ }\bibfield  {title} {\enquote {\bibinfo {title} {Linearly constrained neural networks},}\ }\href@noop {} {\bibfield  {journal} {\bibinfo  {journal} {arXiv preprint arXiv:2002.01600}\ } (\bibinfo {year} {2020})}\BibitemShut {NoStop}%
\bibitem [{\citenamefont {Kingma}\ and\ \citenamefont {Ba}(2014)}]{kingma2014adam}%
  \BibitemOpen
  \bibfield  {author} {\bibinfo {author} {\bibfnamefont {D.~P.}\ \bibnamefont {Kingma}}\ and\ \bibinfo {author} {\bibfnamefont {J.}~\bibnamefont {Ba}},\ }\bibfield  {title} {\enquote {\bibinfo {title} {Adam: A method for stochastic optimization},}\ }\href@noop {} {\bibfield  {journal} {\bibinfo  {journal} {arXiv preprint arXiv:1412.6980}\ } (\bibinfo {year} {2014})}\BibitemShut {NoStop}%
\bibitem [{\citenamefont {Mo}, \citenamefont {Ling},\ and\ \citenamefont {Zeng}(2022)}]{mo2022data}%
  \BibitemOpen
  \bibfield  {author} {\bibinfo {author} {\bibfnamefont {Y.}~\bibnamefont {Mo}}, \bibinfo {author} {\bibfnamefont {L.}~\bibnamefont {Ling}}, \ and\ \bibinfo {author} {\bibfnamefont {D.}~\bibnamefont {Zeng}},\ }\bibfield  {title} {\enquote {\bibinfo {title} {Data-driven vector soliton solutions of coupled nonlinear schr{\"o}dinger equation using a deep learning algorithm},}\ }\href@noop {} {\bibfield  {journal} {\bibinfo  {journal} {Physics Letters A}\ }\textbf {\bibinfo {volume} {421}},\ \bibinfo {pages} {127739} (\bibinfo {year} {2022})}\BibitemShut {NoStop}%
\bibitem [{\citenamefont {Arora}(2015)}]{arora2015optimization}%
  \BibitemOpen
  \bibfield  {author} {\bibinfo {author} {\bibfnamefont {R.~K.}\ \bibnamefont {Arora}},\ }\href@noop {} {\emph {\bibinfo {title} {Optimization: algorithms and applications}}}\ (\bibinfo  {publisher} {CRC press},\ \bibinfo {year} {2015})\BibitemShut {NoStop}%
\bibitem [{\citenamefont {Petersen}\ and\ \citenamefont {Voigtlaender}(2018)}]{petersen2018optimal}%
  \BibitemOpen
  \bibfield  {author} {\bibinfo {author} {\bibfnamefont {P.}~\bibnamefont {Petersen}}\ and\ \bibinfo {author} {\bibfnamefont {F.}~\bibnamefont {Voigtlaender}},\ }\bibfield  {title} {\enquote {\bibinfo {title} {Optimal approximation of piecewise smooth functions using deep relu neural networks},}\ }\href@noop {} {\bibfield  {journal} {\bibinfo  {journal} {Neural Networks}\ }\textbf {\bibinfo {volume} {108}},\ \bibinfo {pages} {296--330} (\bibinfo {year} {2018})}\BibitemShut {NoStop}%
\bibitem [{\citenamefont {Heaton}(2018)}]{heaton2018ian}%
  \BibitemOpen
  \bibfield  {author} {\bibinfo {author} {\bibfnamefont {J.}~\bibnamefont {Heaton}},\ }\bibfield  {title} {\enquote {\bibinfo {title} {Ian goodfellow, yoshua bengio, and aaron courville: Deep learning: The mit press, 2016, 800 pp, isbn: 0262035618},}\ }\href@noop {} {\bibfield  {journal} {\bibinfo  {journal} {Genetic programming and evolvable machines}\ }\textbf {\bibinfo {volume} {19}},\ \bibinfo {pages} {305--307} (\bibinfo {year} {2018})}\BibitemShut {NoStop}%
\bibitem [{\citenamefont {Finlay}\ and\ \citenamefont {Oberman}(2021)}]{finlay2021scaleable}%
  \BibitemOpen
  \bibfield  {author} {\bibinfo {author} {\bibfnamefont {C.}~\bibnamefont {Finlay}}\ and\ \bibinfo {author} {\bibfnamefont {A.~M.}\ \bibnamefont {Oberman}},\ }\bibfield  {title} {\enquote {\bibinfo {title} {Scaleable input gradient regularization for adversarial robustness},}\ }\href@noop {} {\bibfield  {journal} {\bibinfo  {journal} {Machine Learning with Applications}\ }\textbf {\bibinfo {volume} {3}},\ \bibinfo {pages} {100017} (\bibinfo {year} {2021})}\BibitemShut {NoStop}%
\end{thebibliography}%
%

\end{document}